\documentclass[aps,prl,reprint,groupedaddress]{revtex4-1}

\usepackage{graphicx,amsmath,color,amsfonts,bm}

\newcommand{\be}{\begin{equation}}
\newcommand{\ee}{\end{equation}}
\newcommand{\bea}{\begin{eqnarray}}
\newcommand{\eea}{\end{eqnarray}}

\begin{document}

\title{Hydrodynamic sound modes and Galilean symmetry breaking in a magnon fluid}

\author{Joaquin F. Rodriguez-Nieva,$^1$ Daniel Podolsky,$^{2,3}$ and Eugene Demler$^1$}
\affiliation{$^1$Department of Physics, Harvard University, Cambridge, MA 02138, USA}
\affiliation{$^2$Department of Physics, Technion, Haifa 32000, Israel}
\affiliation{$^3$ITAMP, Harvard-Smithsonian Center for Astrophysics, Cambridge, Massachusetts 02138, USA}

\date{\today}
\begin{abstract}

  The non-interacting magnon gas description in ferromagnets breaks down at finite magnon density where momentum-conserving collisions between magnons become important. Observation of the collision-dominated regime, however, has been hampered by the lack of probes to access the energy and lengthscales characteristic of this regime. Here we identify a key signature of the collision-dominated hydrodynamic regime---a magnon sound mode---which governs dynamics at low frequencies and can be detected with recently-introduced spin qubit magnetometers. The magnon sound mode is an excitation of the longitudinal spin component with frequencies below the spin wave continuum in gapped ferromagnets. We also show that, in the presence of exchange interactions with SU(2) symmetry, the ferromagnet hosts an usual hydrodynamic regime that lacks Galilean symmetry at all energy and lengthscales. The hydrodynamic sound mode, if detected, can lead to a new platform to explore hydrodynamic behavior in quantum materials. 
  
\end{abstract}


\maketitle

{\bf Introduction.}---The presence of symmetries and conservation laws can affect the universal dynamics of interacting quantum systems in dramatic ways. One example is the recently observed hydrodynamic regime in graphene where, in a wide range of temperatures, fast momentum-conserving collisions lead to viscous electron transport \cite{2015polini,2016bandurin,2016crossno,2016levitovfalkovich,2017superballisticconduction,2017superballisticflowexp}. This unusual electron transport behavior, also proposed in a variety of other quantum systems\cite{2016superconductivity,2016lucassoundmodes,2016lucas-weyl,integrable1,2017integrability2,2017resistivitybound,2017curvedspace,2017lucadensitywave,2017lucahallviscosity,2017moorehydro}, differs from the more conventional ballistic and diffusive regimes. The giant leap in our understanding of quantum transport that resulted from the study of hydrodynamics in graphene motivates us to raise two new questions: (i) are there other experimental platforms beyond graphene that enable us to probe new regimes of hydrodynamics in quantum materials? (ii) can additional symmetries give rise to qualitatively distinct transport features? 

Here we show that a magnon gas describing low-energy excitations in a Heisenberg ferromagnet can enter an unusual hydrodynamic regime in a wide range of temperatures and frequencies when SU(2) symmetry is present, and we propose an experimental protocol to detect hydrodynamic modes using spin qubit magnetometers \cite{2014nvreview,2017cappellaro}. As we argue below, the description of long wavelength excitations in terms of ballistic spin waves, or magnons, relies on a vanishingly small magnon-magnon interaction strength which renders relaxation processes at the bottom of the band very inefficient. However, as temperature increases and the thermal magnon population occupies larger momentum states, {\it momentum conserving} collisions give rise to a relaxation length $\ell$ which steeply decreases with temperature $T$ and magnon density $n$\cite{1956dyson}: 
\be
{\ell} = \frac{1}{na^{d-1}}\left(\frac{J}{T}\right)^{\frac{d+1}{2}}.
\label{eq:relaxationlength}
\ee
Here $a$ is the lattice spacing, $J$ is the exchange coupling, and $d\ge 2$ is the system's dimension. For an intermediate temperature range such that Umklapp scattering can be neglected ($T \ll J$), but large enough such that $\ell \ll L$ for the characteristic length $L$ of the system, hydrodynamic behavior emerges. For instance, for moderately small occupation numbers ($na^d \sim 0.1$) and temperature below the Curie temperature ($T/J \sim 0.2$), $\ell \sim 50\,{\rm nm}$ is much smaller than a typical sample length $L\sim 10\,\mu{\rm m}$ (here we used $a = 0.5\,{\rm nm}$ and $d=2$). 

\begin{figure}[b]
  \centering\includegraphics[scale=1.0]{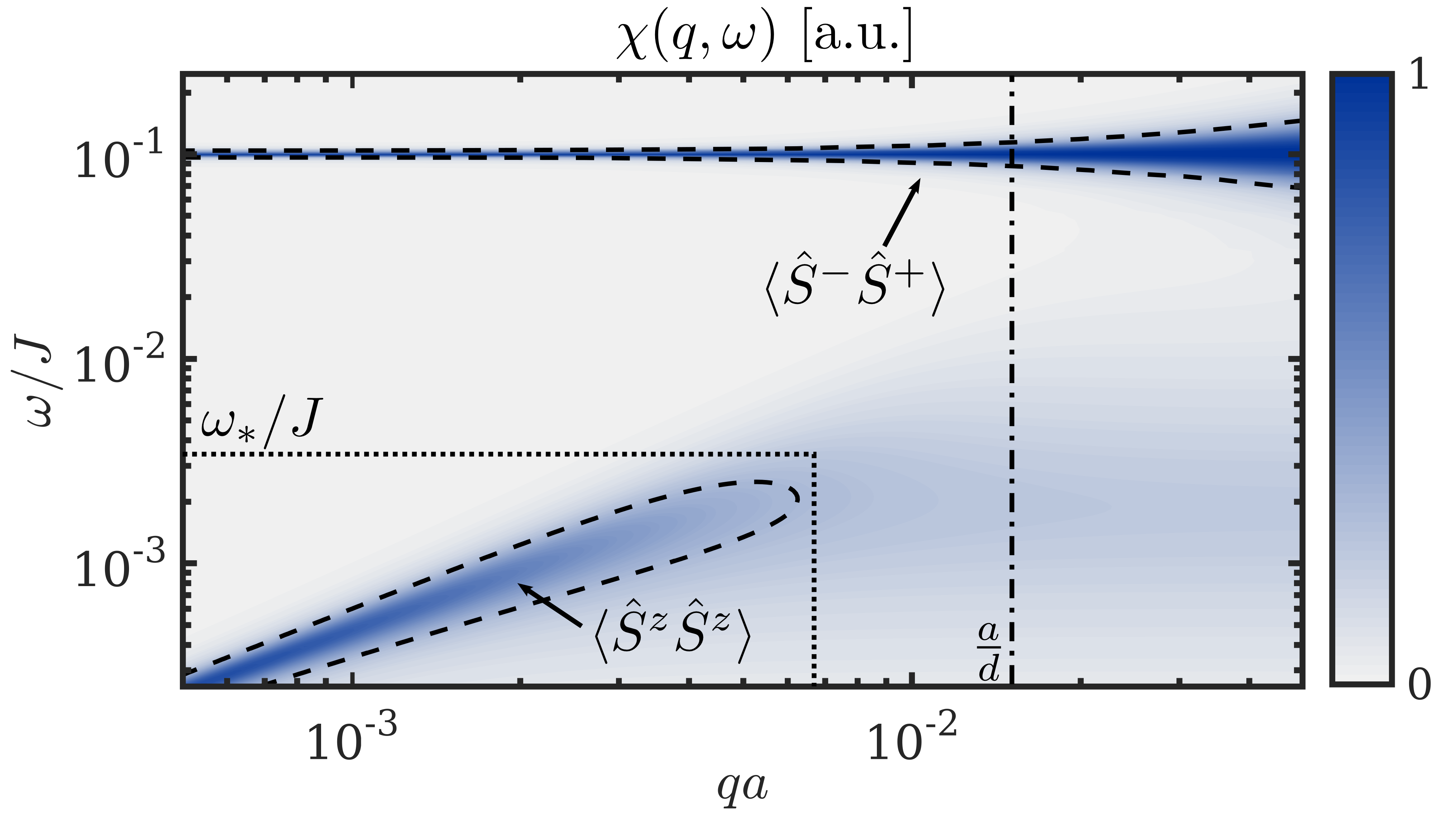}
\vspace{-6mm}
\caption{Spectral function $\chi(q,\omega)=\chi_{+-}(q,\omega)+\chi_{-+}(q,\omega)+4 \chi_{zz}(q,\omega)$ exhibiting single magnon excitations at the Zeeman energy $\omega = \Delta $, induced by a finite $\langle S^-S^+\rangle$, and a linearly dispersing sound mode at low frequencies induced by magnon density fluctuations, $\langle \hat{S}^z \hat{S}^z \rangle$. The sound mode is damped above frequencies $\omega_*$ by viscous forces.}
\label{fig:spectrum}
\end{figure} 

A key signature of momentum-conserving collisions is the existence of a sound mode. As shown in Fig.\ref{fig:spectrum}, the sound mode is manifested as an excitation of the longitudinal spin correlator, $\langle \hat{S}^z \hat{S}^z\rangle$, where $\hat{S}^z$ is related to the magnon density ${n}$ via $\langle \hat{S}^z \rangle = S(1-{n}a^2)$, and is analogous to a second sound in a superfluid. As a result, spin fluctuation measurements can provide clear-cut signatures of the sound mode, as shown below. We also distinguish magnon hydrodynamics from hydrodynamics in electron fluids where, rather than sound modes, the system hosts plasmon modes; this qualitatively distinct behavior arises because longitudinal charge fluctuations are mediated by long-ranged Coulomb interactions\cite{2018electronsound}.

A second signature of the hydrodynamic regime in a magnon fluid is that collisions between quasiparticles is strongly constrained by SU(2) symmetry and gives rise to strong momentum dependence of the magnon-magnon interaction, see Eq.(\ref{eq:Heffective}) below. This feature has important consequences for universal dynamics. First, Galilean symmetry is {\it intrinsically} broken by the interaction at all length and energy scales, and differs from usual hydrodynamics in lattice systems where Galilean symmetry is broken {\it only} at energy scales comparable to the single-particle bandwidth ({\it i.e.}, when deviations from quadratic dispersion are sizable). While Galilean symmetry breaking can also be induced by non-linearities in planar ferromagnets in the presence of a spin texture\cite{2017galileansymmetrybreaking,2011demler,2017demler}, here we obtain Galilean symmetry breaking even in a {\it stationary} fluid without a spin texture. Second, vanishingly small scattering matrix elements suppress collisions between magnons and the condensate that arises due to symmetry breaking. Such suppresion justifies why the dispersion of magnons---the Goldstone modes of the ferromagnet---remain quadratic in the symmetry-broken phase, contrary to U(1)-symmetry breaking where interactions between quasiparticles and the condensate renormalize the quasiparticle dispersion ({\it i.e.}, first sound) and where a `two-fluid' hydrodynamic description is necessary. 

Previous works on hydrodynamics in ferromagnets assume momentum relaxation due to Umklapp scattering ($T\approx J$) or disorder, as first described by Halperin and Hohenberg \cite{1969halperinhohenberg}. Such momentum-relaxing effects give rise to diffusive particle and energy transport. Although a few authors \cite{1968reiter,1969michel,1970Schwabl} made the case for momentum-conserving hydrodynamic behavior in a magnon gas, no experimental signature of this regime has been observed to date. Arguably, the energy scales ($\sim$meV) and wavevectors ($\gtrsim 1/a$) accessible by neutron scattering, the main probe of ferromagnets at the time, were too large to access the low frequency, long-wavelength regime in which hydrodynamic sound modes live. 

We argue that recent experiments \cite{2015nvmagnons,2017du} have opened new pathways to observe and study hydrodynamic behavior in spin systems. First, ultraclean ferromagnetic materials, such as yttrium iron garnet (YIG), allow ballistic propagation of magnons in macroscopic scales without scattering by impurities and phonons. Second, {\it independent} control of temperature and chemical potential is now possible via a combination of heating and driving and, therefore, enables us to explore all possible regimes from non-interacting magnon gases to interacting magnon fluids. Finally, magnetic spectroscopy with spin qubits allows to access spin fluctuations at the energy and lengthscales relevant for hydrodynamics. Besides spin waves \cite{2015nvmagnons,2017du}, such probes have been used to image single spins \cite{2013singlespinimaging}, domain walls \cite{2016nvdomainwalls}, and to study electron transport in metals \cite{2015kolkowitz}. The have also been proposed to access the hydrodynamic regime in graphene \cite{2017kartieknoise} and one-dimensional systems \cite{2018nv-wire}, to study magnon consensation in ferromagnets \cite{2018flebus}, and to diagnose ground states in frustrated magnets\cite{2018nv-sl}. 

\vspace{1mm}{\bf Microscopic model.}---We consider a two-dimensional Heisenberg ferromagnet in the presence of a Zeeman field and a small exchange anisotropy $\epsilon > 0$: 
\be
\hat{H} = -J\sum_{\langle jj'\rangle}\left({\hat{\bm S}}_j \cdot {\hat{\bm S}}_{j'} + \epsilon{\hat{S}}_j^z \cdot {\hat{S}}_{j'}^z\right) + \Delta \sum_j \hat{S}_j^z.
\label{eq:Hamiltonian}
\ee
Here $j$ labels the lattice site, $\sum_{\langle j j' \rangle}$ denotes summation over nearest neighbors, and we take periodic boundary conditions in each spatial direction. We assume that the spin system has $N$ lattice sites on a square lattice, each containing a spin $S$ degree of freedom which satisfies the commutation relations $[{\hat S}_j^z,{\hat S}_{j'}^\pm] = \pm \delta_{jj'}{\hat S}_j^\pm$ and $[{\hat S}_j^+,{\hat S}_{j'}^-] = 2 \delta_{jj'} {\hat S}_j^z$, with ${\hat S}_{j}^\pm = {\hat S}_j^x \pm i {\hat S}_j^y$ the raising and lowering spin operators. The Zeeman term plays an important role experimentally because it allows to separate the magnon continuum from the gapless sound mode. 

With the objective of deriving an effective model describing the low energy manifold of $\hat{H}$, we recall that one magnon states $|{\bm k}\rangle = \hat{S}_{\bm k}^+|{\rm F}\rangle$, with $|{\rm F}\rangle = |\downarrow\downarrow\ldots\downarrow\rangle$ denoting the ferromagnetic ground state and $\hat{S}_{\bm k}^+ =\frac{1}{\sqrt{N}} \sum_j e^{-i{\bm k}\cdot {\bm r}_{j}}\hat{S}_{j}^+$, are exact eigenstates of $\hat{ H}$ with energies 
\be 
\varepsilon_{\bm k} = \Delta + JS[\phi_0 (1+\epsilon)- \phi_{\bm k}], \quad \phi_{\bm k} = \sum_{{\boldsymbol\tau} \in {\rm NN}} e^{i{\bm k}\cdot {\boldsymbol\tau}}.
\label{eq:energy} 
\ee
Two magnon states $|{\bm k},{\bm p}\rangle = \frac{1}{2S}\hat{S}_{\bm k}^+\hat{S}_{\bm p}^+|{\rm F}\rangle$, however, are {\it not} eigenstates of $\hat{H}$\cite{twomagnons,1956dyson,mattisbook}. Indeed, it is straightforward to show that
\be
\begin{array}{c}
\displaystyle\hat{H} | {\bm k},{\bm p}\rangle = (\varepsilon_{\bm k}+\varepsilon_{\bm p})|{\bm k},{\bm p}\rangle + \frac{1}{N}\sum_{\bm q}g_{\bm k,\bm p,\bm q}|{\bm k + \bm q},{\bm p - \bm q}\rangle,\\
\displaystyle g_{\bm k, \bm p,\bm q} = - J\left(\epsilon -\phi_{\bm q} + \phi_{\bm q - \bm p} + \phi_{\bm q+\bm k}-\phi_{\bm k+\bm q-\bm p}\right), 
\end{array}
\label{eq:interaction}
\ee
such that one magnon states are coupled via momentum-conserving collision $g_{\bm k,\bm p,\bm q}$. More generally, an $N$-magnon state ${\hat H}|N\rangle ={\hat H}\left[ \frac{1}{(2S)^{N/2}} \prod_{i}^N \hat{S}_{{\bm k}_i}^+\right]|F\rangle$ can also be decomposed into a diagonal component comprised of individual spin wave energies, and an off-diagonal component containg all possible combinations of two-body collisions \cite{supplement}. When the incoming magnons are close to the bottom of the band, the collision term is approximately $g_{\bm k,\bm p,\bm q}\approx - Ja^2(\epsilon+{\bm k}\cdot{\bm p})$. Importantly, 
whereas collisions between quasiparticles are hardcore in the easy axis ferromagnet, collisions are soft under SU(2) symmetry($\epsilon=0$). We will focus on the latter regime which has remained unexplored (for a discussion on the easy plane ferromagnet with broken U(1) symmetry, see Ref.\,\cite{2016flebus}).

An effective description of the interacting magnon fluid which captures all the features of the parent SU(2) symmetric Hamiltonian in Eq.\,(\ref{eq:Hamiltonian}) is given by
\be
\hat{H} = \int_{\bm x} \frac{\partial_\alpha\hat{\psi}_{\bm x}^\dagger \partial_\alpha\hat{\psi}_{\bm x}}{2m_0}+\frac{Ja^2}{4}\left(\hat{\psi}_{\bm x}^\dagger\hat{\psi}_{\bm x}^\dagger\partial_\alpha\hat{\psi}_{\bm x}\partial_\alpha\hat{\psi}_{\bm x}+{\it h.c.}\right),
\label{eq:Heffective}
\ee
where $m_0=1/2SJa^2$ is the magnon mass and $\hat{\psi}$ is a bosonic operator defined after a Holstein-Primakoff transformation ($\hat{S}_{\bm x}^- \approx \sqrt{2S}\hat{\psi}_{\bm x}$ and $\hat{S}_{\bm x}^+ \approx \sqrt{2S}\hat{\psi}_{\bm x}$), and summation over repeated indices is assumed. Equation (\ref{eq:Heffective}) is valid in the dilute limit $na^d \ll 1$ and small temperature $T\ll J$ such that only small momentum states are occupied.

\vspace{1mm}{\bf Magnon Hydrodynamics without Galilean symmetry.}---The conserved quantities in Eq.(\ref{eq:Heffective}) are $\hat{N} = \int_{\bm x} \hat{n}_{\bm x} = \int_{\bm x} \hat{\psi}_{\bm x}^\dagger\hat{\psi}_{\bm x}$, $\hat{P}_{\alpha} = \int_{\bm x}\hat{p}_{\alpha,\bm x} = \frac{-i}{2}\int_{\bm x}\hat{\psi}_{\bm x}^\dagger \partial_\alpha\hat{\psi}_{\bm x}-{h.c.}$, and $\hat{ H} $. Although $\hat{P}_{\alpha}$ is not strictly conserved in the lattice model (\ref{eq:Hamiltonian}), it becomes conserved in the long-wavelength effective theory after neglecting Umklapp scattering. We use the  {\it local equilibrium approximation} to describe the density matrix as $\hat{\rho} = \prod_{\bm x} \hat{\rho}_{\bm x}$, where space is coarse-gained into regions of size $\ell$. The local density matrix is $\hat{\rho}_{\bm x} = {\rm exp}\left( - \hat{H}/T - u_{\alpha}\hat{P}_{\alpha} - \mu \hat{N}\right)_{\bm x}$, with $(T,u_\alpha,\mu)_{\bm x}$ the position and time-dependent thermodynamic potentials. To compute expectation values, we use a Gaussian approximation of the distribution function $\rho_{\bm x}$ which can be formally implemented by using $\hat{H}/T \approx \sum_{\bm k}\frac{{\bm k}^2}{2{m}T}\hat{\psi}_{\bm k}^\dagger\hat{\psi}_{\bm k}$, with ${m}$ the renormalized magnon mass. As such, any $N$-point correlation functions can be expressed as products of two point correlation functions. Because corrections to the bare mass are small, $\delta m = m - m_0 \sim {\cal O}(na^2T/J)\ll 1$, below we will use $m $ and $ m_0$ interchangeably. The expectation value of conserved quantities ($\langle \hat{N}\rangle_{\bm x} = n$, $\langle \hat{P}_\alpha\rangle_{\bm x} = n p_\alpha $, $\langle \hat{H}\rangle_{\bm x} = n\varepsilon  $) are given by: 
\be
{n} = \frac{mT}{2\pi}g_1(z),\quad\quad p_\alpha= mu_\alpha,\quad\quad \theta = \frac{Tg_2(z)}{g_1(z)}, 
\label{eq:qerelations}
\ee
where the thermal energy $\theta$ is related to energy density via $\varepsilon = \theta + \frac{(1-na^2/4S)p^2}{2m}$. In Eq.(\ref{eq:qerelations}), $z = e^{-\mu/T}$ is the fugacity, and $g_{\nu}(z)$ is the Bose integral, $g_\nu(z) = \frac{1}{\Gamma(\nu)}\int_0^\infty  \frac{dy y^{\nu-1}}{e^y/z-1}$[$\Gamma(\nu)$: Gamma function]. 

One crucial aspect of Eq.(\ref{eq:Heffective}) is that the particle current operator, defined as $\partial_\alpha \hat{J}_{\alpha} = -i[\hat{ H},\hat{n}_{\bm x}]$, is {\it not equal} to $\hat{P}_{\alpha}$; instead, $\hat{J}_{\alpha}$ takes the form $\hat{J}_{\alpha} = \hat{P}_{\alpha}/m_0 + \frac{iJa^2}{2}(\hat{\psi}_{\bm x}^\dagger\hat{\psi}_{\bm x}^\dagger\hat{\psi}_{\bm x}\partial_\alpha\hat{\psi}_{\bm x} - {h.c.})$, and gives rise to Galilean symmetry breaking. 
For the purposes of this work, the main consequence of Galilean symmetry breaking is that conserved quantities flow with a drift velocity $v_{\alpha} = \langle \hat{J}_{\alpha}\rangle /n$ which is different from the thermodynamic potential $u_\alpha$: 
\be
v_{\alpha} = (1 - \gamma) u_{\alpha}, \quad \gamma = \frac{na^2}{S}.
\ee
In particular, as shown in the Supplement, the particle current $J_\alpha$, the momentum current $\Pi_{\alpha\beta}=\langle\hat{\Pi}_{\alpha\beta}\rangle $, and the energy current $Q_\alpha = \langle \hat{Q}_\alpha\rangle$, are given by 
\be
\begin{array}{rl}
 \displaystyle J_{\alpha} & \displaystyle = nv_{\alpha}, \\ 

\displaystyle \Pi_{\alpha\beta} & \displaystyle = np_{\alpha}v_{\beta} + P_{\alpha\beta}, \\ 
\displaystyle Q_{\alpha} & \displaystyle = n\varepsilon v_{\alpha}  + P_{\alpha\beta}v_\beta+ q_\alpha.
\end{array}
\label{eq:currents}
\ee 
Here $P_{\alpha\beta} = (n\theta-\frac{\gamma n p^2}{2m})\delta_{\alpha\beta}+\tilde{P}_{\alpha\beta}$ is the pressure tensor, with $\tilde{P}_{\alpha\beta}$ the dissipative (viscous) component, and $q_\alpha$ is the heat current (both $\tilde{P}_{\alpha\beta}$ and $q_\alpha$ will be defined explicitly below). The continuity equations for each of the conserved charges lead to the hydrodynamic equations:
\be
\begin{array}{l}
\displaystyle \dot{n} + \partial_\alpha (n{v}_\alpha) = 0, 
\\ 
\displaystyle  \dot{p}_\alpha + v_\beta \partial_\beta p_\alpha = - \frac{1}{n}\partial_\beta P_{\alpha\beta},
\\ 
\displaystyle \dot{\theta} + v_\alpha \partial_\alpha \theta = -\frac{1}{n}\partial_\alpha q_\alpha - \frac{1}{n}P_{\alpha\beta}\partial_\alpha v_\beta-\gamma\frac{p^2}{2m}\partial_\alpha v_\alpha,
\end{array}
\label{eq:hydro}
\ee
which resemble usual hydrodynamic equations for a classical fluid with the caveat that convective terms contain $v_{\alpha}$ rather than $u_\alpha$. We recall that the `single fluid' equations (\ref{eq:hydro}) do not include dynamics of the condensate due to the zero coupling with ${\bm k}=0$ modes in the SU(2) symmetric Hamiltonian. Although here we will focus mainly on a stationary fluid ($v_\alpha\approx 0$), Galilean symmetry breaking gives rise to a variety of interesting effects at finite velocities, including velocity-dependent transport coefficients, anomalous viscous terms, and anisotropic dispersion of hydrodynamic fluctuations, to name a few. 

\vspace{1mm}{\bf Dissipative effects}---We incorporate dissipation effects phenomenologically using the relaxation time approximation, see Supplement. This approximation allows us to relate the non-equilibrium magnon distribution to gradients in $\eta_j = (n,u_\alpha,\theta)$, {\it i.e.} $n_{\bm k} = \bar{n}_{\bm k} + \tau_{\bm k}\sum_j (\partial \bar{n}_{\bm k}/\partial \eta_j)(\partial_t+{\bm v}_{\bm k}\cdot\nabla_{\bm r})\eta_j$, where $\tau_{\bm k}$ is a momentum-dependent relaxation time (see details in Supplement). As a result, $\tilde{P}_{\alpha\beta}$ and $q_\alpha$ can be written $\tilde{P}_{\alpha\beta} = \mu\left(\partial_\alpha u_\beta + \partial_\beta u_\alpha - \delta_{\alpha\beta}\partial_\gamma u_\gamma \right)$, and $q_\alpha = \kappa_n \partial_\alpha n + \kappa_\theta \partial_{\alpha}\theta$. For a two-dimensional magnon gas with quadratic dispersion and collision rate of the form $1/\tau_{\bm k}\propto {\bm k}^2$, we find that, within the relaxation time approximation, dissipation is dominated by the viscous effects $\mu \sim \frac{J^2}{T}$. While we will keep track of $\kappa_n$ and $\kappa_\theta$ in our equations of motion, we set $\kappa_n = \kappa_\theta = 0$ in the numerics.

\vspace{1mm}{\bf Hydrodynamic modes.}---The central result of our work is the emergence of a collective hydrodynamic mode in a spin system, which is directly accessible with noise magnetometry. This mode originates from the longitudinal spin fluctuations which can be quantified by the retarded correlator 
\be
\chi_{zz}({\bm q},\omega) = -i \int_{0}^{\infty} dt e^{i\omega t}\sum_\tau e^{-i{\bm q}\cdot{\boldsymbol\tau}}\langle [\hat{S}_{i}^z(t),\hat{S}_{i+\tau}^z(0)]\rangle.
\ee
This is equivalent to computing density fluctuation because $\hat{S}_i^z = -S(1-\hat{n}_i)$. With this objective in mind, we first linearize 
Eq.(\ref{eq:hydro}) around the equilibrium values, $n(\bm r,t) = \bar{n} + \delta n(\bm r,t)$, $\theta(\bm r,t) = \bar\theta + \delta \theta(\bm r ,t)$, and $v_\alpha(\bm r,t) = \delta v_\alpha(\bm r,t)$, and go to momentum space: 
\be
\begin{array}{c}
\mathbb{A} \left(\begin{array}{c}
\delta n \\ 
\delta v_\parallel \\
\delta \theta
\end{array}\right)  = 
\left(\begin{array}{c}
0 \\ 
iF_\parallel/m \\
0
\end{array}\right),\\
\mathbb{A} = \left(\begin{array}{ccc}
\omega & - \bar{n} q & 0 \\
-\bar\theta q/{m \bar n} & \omega/(1-\gamma) + i\mu q^2/\bar n & - q/m \\
-i\kappa_nq^2/\bar{n} & - \bar{\theta} q &  \omega - i\kappa_\theta q^2/\bar{n}
\end{array}\right).
\end{array}
\label{eq:hydrolinear}
\ee
The coupling between $\delta n$, $\delta v_\parallel$ and $\delta\theta$ gives rise to two propagating modes and one diffusive mode. The transverse momentum component, $\delta u_\perp$, which does not couple to $\delta n$, gives rise to an extra diffusive mode, $(\omega + i \mu q^2/\bar{n})\delta v_\perp = iF_\perp/m$. Here we included in our equations a fluctuating parallel (transverse) force $F_\parallel$ ($F_\perp$). Close to thermal equilibrium, the density-density correlation function can be obtained from Eq.(\ref{eq:hydrolinear}) using the fluctuation-dissipation theorem: 
\be
\chi_{zz}(q,\omega) = \frac{JS^2(\bar{n}qa^2)^2}{\omega^2/(1-\gamma) - \zeta(q,\omega)\bar{\theta}q^2/m + i \mu\omega q^2 / \bar{n}}, 
\ee
where $\zeta (q,\omega) = 1+\frac{\omega - i\kappa_n q^2/\bar{\theta}}{\omega + i\kappa_\theta q^2/\bar{n}}\approx 2$ at the intermediate/large frequency range of interest. In this regime, the response function exhibits a linearly dispersing sound mode $\omega = v_{\rm s}q$, with $\quad v_{\rm s} = a\sqrt{2(1-\gamma)J\theta}$, see Fig.\ref{fig:spectrum}. 

\begin{figure}
  \centering\includegraphics[scale=1.0]{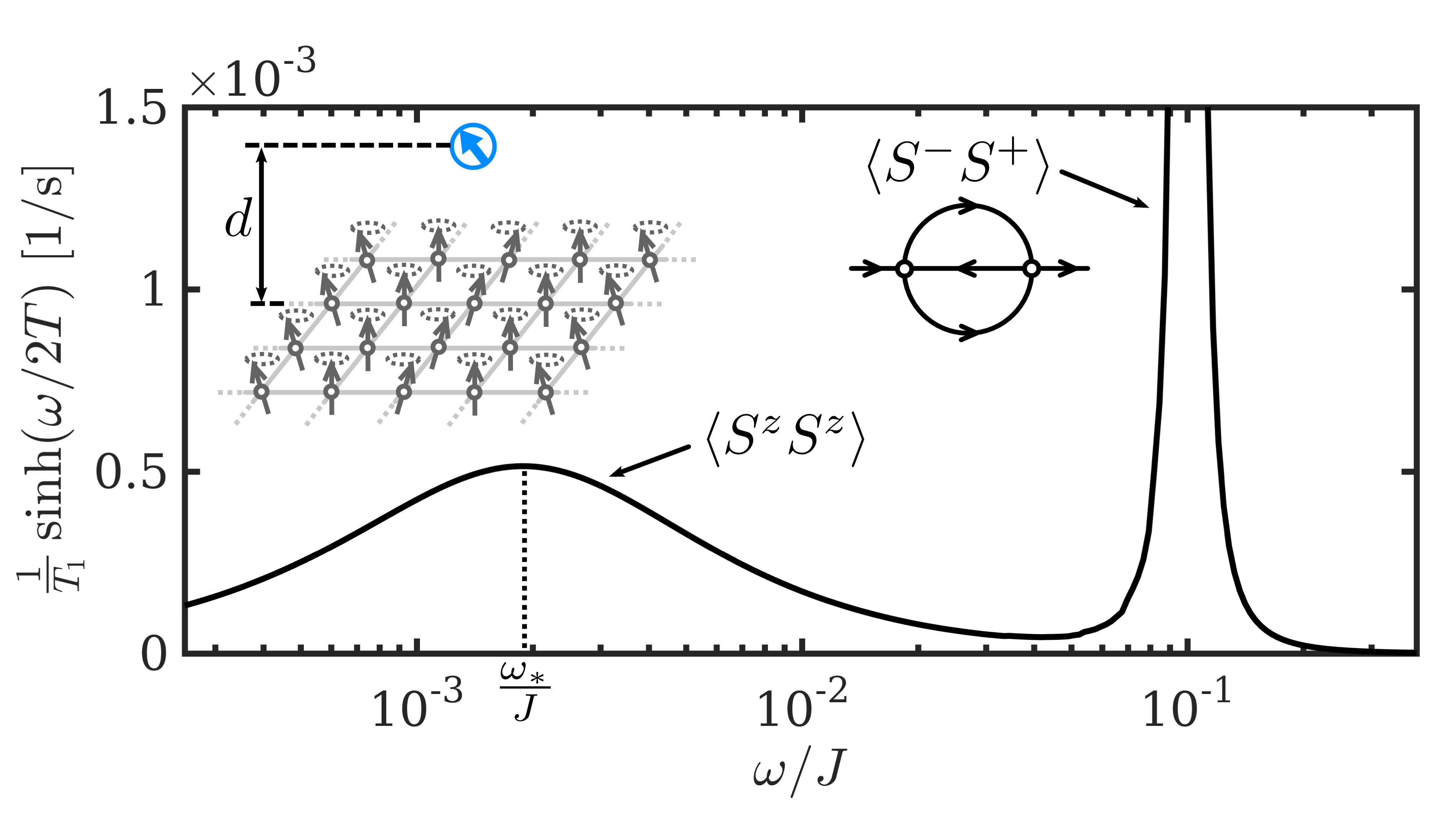}
\vspace{-6mm}
  \caption{Relaxation time [normalized by $\sinh(\omega/2T)$] of a spin qubit located a distance $d$ from the 2D ferromagnet. Besides the characteristically large relaxation rate induced by spin relaxation due to emission of spin waves at energy $\Delta$, the relaxation rate exhibits a peak below the ferromagentic gap induced by emission of sound modes with velocity $v_{\rm s}$. Parameters used: $na^2 = 0.03$, $T/J =0.2$, $\Delta/J = 0.1$, $a = 0.3\,{\rm nm}$, and $d = 20\,{\rm nm}$.}
\label{fig:relaxationtime}
\end{figure}

\vspace{1mm}{\bf Detection of the sound mode.}---We consider a spin-1/2 qubit with an intrinsic level splitting $\omega$ placed a distance $d$ above the magnetic insulator. The combined dynamics of the qubit and ferromagnet is governed by the Hamiltonian $\hat{H}_{\rm total} = \hat{H} + \hat{H}_{\rm c} + \hat{H}_{\rm q}$, where $\hat{ H}_{\rm q}$ is the spin qubit Hamiltonian $\hat{H}_{\rm q} = \frac{1}{2}\omega \sigma_z$ with polarizing field assumed to be aligned in the $z$ direction. The term $\hat{H}_{\rm c}$ is the qubit-ferromagnet coupling induced by dipole-dipole interactions:
\be
\hat{ H}_{\rm c} = \frac{\mu_{\rm B}^2}{2}\hat{\boldsymbol\sigma}\cdot\hat{\bm B}, \quad \hat{\bm B} =\frac{1}{4\pi} \sum_j \left[\frac{\hat{\bm S}_j}{r_j^3} - \frac{3(\hat{\bm S}_j\cdot{\bm r}_j) {\bm r}_j}{r_j^5}\right],
\label{eq:HqF}
\ee
where ${\bm r}_j = (x_j,y_j,-d)$ is the relative position between the $i$-th spin in the 2D lattice and probe. The relaxation time of the spin qubit can be obtained from Fermi Golden's rule $1/T_1 = \frac{\mu_{\rm B}^2}{2}\int_{-\infty}^{\infty} dt e^{i\omega t}\langle \{ \hat{B}^-(t), \hat{B}^+(0)\}\rangle$, where $\{,\}$ denotes anticommutation\cite{supplement}. Replacing Eq.(\ref{eq:HqF}) into $1/T_1$ and using the fluctuation-dissipation theorem, the relaxation time can be expressed in terms of spin correlation functions: 
\be
\begin{array}{rl}
\displaystyle \frac{1}{T_1} = & \displaystyle {\rm coth}\left(\frac{\omega}{2T}\right) \frac{\mu_{\rm B}^2}{2a^2} \int \frac{d^2{\bm q}}{(2\pi)^2} e^{-2|{\bm q}|d} |{\bm q}|^2 \left[\chi_{-+}''({\bm q},\omega) \right. \\ & \\
& \displaystyle\left. +\chi_{+-}''({\bm q},\omega)+4\chi_{zz}''({\bm q},\omega)\right], 
\end{array}
\label{eq:relaxationtime}
\ee
where we denote $\chi_{\alpha\beta}'' = - {\rm Im}[\chi_{\alpha\beta}]$, and $\chi_{\alpha\beta}^{\rm R}({\bm q},\omega) = -i\int_0^\infty dt \langle[\hat{S}_{-\bm q}^\alpha(t),\hat{S}_{\bm q}^\beta(0)]\rangle$. Figure\,\ref{fig:spectrum} shows the integrand of Eq.(\ref{eq:relaxationtime}), and Fig.\ref{fig:relaxationtime} shows the spin relaxation time as a function of $\omega$ induced by longitudinal and transverse spin fluctuations (we assumed a constant magnon population $\bar{n}$ and $T$). The correlators $\chi_{\pm\mp}({\bm q},\omega)$ are related to single-magnon production/absorption, which we assume to be given by $\chi_{+-}^{-1}({\bm q},\omega) = \omega-\omega_{\bm q}+i\Sigma''({\bm q},\omega)$, where $\Sigma''({\bm q},\omega)\sim \frac{T\omega}{J}(qa)^2$ (valid for $z\sim 1$ and $\omega \ll T$) is the imaginary part of the self-energy computed from the bubble diagram, see inset of Fig.\ref{fig:relaxationtime} and details in the Supplement. We also note that, in Fig.\ref{fig:relaxationtime}, we normalize $1/T_1$ with $\coth(\omega /2 T)$ to capture the spectral contribution of spin fluctuations rather than its amplitude. Figure\,\ref{fig:relaxationtime} is the main result of this work, and shows a clear fingerprint of the sound mode within the gap of the ferromagnet. 

\vspace{1mm}{\bf Dipolar interactions.}---Contrary to classical and electron fluids wherein particles cannot be created or annhiliated, conservations laws are not as robust in a magnon fluid and, therefore, should be subject to scrutiny. Dipolar interactions lead to magnon decay via three-magnon processes, particularly in thin layers with a canted ferromagnetic order parameter. Assuming a magnon distribution with $z<1$, we estimate the typical magnon decay time induced by a dipolar term $\hat{H}_{\rm d} = \frac{g_{\rm d}}{2} \sum_{jj'}\left[\frac{\hat{\bm S}_j\cdot\hat{\bm S}_{j'}}{r_{jj'}^3}- \frac{3(\hat{\bm S}_j\cdot{\bm r}_{jj'})(\hat{\bm S}_{j'}\cdot{\bm r}_{jj'})}{r_{jj'}^5}\right]$, with $g_{\rm d} = \mu_{\rm B}^2/4\pi$. As shown in the Supplement, this gives values on the ballpark $\frac{1}{\bar{n}}\frac{d\bar{n}}{dt} \sim \frac{g_{\rm d}^2}{J}(z^2-z^3)\sim {\rm MHz}$, several orders of magnitude smaller than the typical GHz frequencies that typical spin-qubit magnetometers can access. As a result, sound modes are expected to be well defined excitations in a wide range of frequencies, from MHz to several GHz.

\vspace{1mm}{\bf Conclusion.}---Our predictions, which can be tested in ongoing experiments using spin qubit magnetometers on ferromagnetic insulators, provide distinct signatures of hydrodynamic behavior in spin systems. Although the sound mode is its most distinctive feature, the strong momentum dependence of the magnon-magnon interaction induced by the SU(2) symmetry suggests that ferromagnets can also host anomalous transport not achievable in classical and electron fluids. 

\section{Acknowledgements.}

We acknowledge useful discussions with D. Abanin, T. Andersen, S. Chernyshev, C. Du, M.~D. Lukin, A. Rosch, D. Sels, A. Yacoby, J. Sanchez-Yamagishi, and T. Zhou. JFRN and ED acknowledge support from Harvard-MIT CUA, AFOSR-MURI: Photonic Quantum Matter (award FA95501610323), DARPA DRINQS program (award D18AC00014). DP thanks support by the Joint UGS-ISF Research Grant Program under grant number 1903/14 and by the National Science Foundation through a grant to ITAMP at the Harvard-Smithsonian Center for Astrophysics.

%


\clearpage

\renewcommand{\thefigure}{S\arabic{figure}}
\renewcommand{\theequation}{S\arabic{equation}}
\setcounter{page}{1}
\setcounter{equation}{0}
\setcounter{figure}{0}

\begin{widetext}

\begin{center}
{\large\bf Supplement for `Hydrodynamic sound modes and Galilean symmetry breaking in a magnon fluid'}

\vspace{4mm}

Joaquin F. Rodriguez-Nieva,$^1$ Daniel Podolsky,$^{2,3}$ and Eugene Demler$^1$

{\small\it $^1$Department of Physics, Harvard University, Cambridge, MA 02138, USA}

{\small\it $^2$Department of Physics, Technion, Haifa 32000, Israel}

{\small\it $^3$ITAMP, Harvard-Smithsonian Center for Astrophysics, Cambridge, Massachusetts 02138, USA}

\end{center}


\vspace{6mm}

The outline of the Supplement is as follows. In Sec.\,A, we present a derivation of the two magnon collision term in the Heisenberg model. In Sec.\,B, we numerically evaluate the magnon relaxation time due to exchange coupling. In Sec.\,C, we derive the current operators and compute their expectation value using the local equilibrium and gaussian approximation of the distribution function. In Sec.\,D, we compute the viscosity of the magnon fluid using the relaxation time approximation. In Sec.\,E, we evaluate the sunrise diagram which gives rise to a finite linewidth to the single magnon emission/absorption process. In Sec.\,F, we provide the computational steps to obtain Eq.(\ref{eq:relaxationtime}) of the main text. In Sec.\,G, we provide a detailed discussion of dipole-dipole interactions and estimate the typical magnon leakage rate. 

\section{A. Eigenstates of the Heisenberg ferromagnet}

To show the origin of the collision term in Eq.(\ref{eq:interaction}), here we calculate the eigenstates of the Heisenberg Hamiltonian for increasing magnon number. The discussion closely follows that in Ref.\,\cite{mattisbook}. Given that $[\hat{ H},\sum_iS_i^z]=0$, we can label eigenstates with the total number of spin flips. Before computing the eigenstates, it is useful to first write the Hamiltonian in momentum space, 
\be
\hat{ H} = - \frac{J}{4} \sum_{\bm k}\gamma_{\bm k} \left[\hat{S}_{-\bm k}^+\hat{S}_{\bm k}^- + \hat{S}_{-\bm k}^-\hat{S}_{\bm k}^+ + 2 {\hat S}_{-\bm k}^z{\hat S}_{\bm k}^z\right] + \Delta\sum_{j}S_{j}^z,\quad \gamma_{\bm k} = \sum_{\boldsymbol\tau}e^{i{\bm k}\cdot{\boldsymbol\tau}},
\ee
where ${\boldsymbol\tau}$ labels the four nearest neighbor vectors. The spin operators in momentum space satisfy the commutation relations $\left[ {\hat S}_{\bm k}^z , {\hat S}_{{\bm k}'}^\pm\right] = \pm \frac{{\hat S}_{{\bm k}+{\bm k}'}^\pm}{\sqrt{N}}$, $\left[ {\hat S}_{\bm k}^+ , {\hat S}_{{\bm k}'}^-\right] = \frac{2 {\hat S}_{{\bm k}+{\bm k}'}^z}{\sqrt{N}}$. 

\subsection{Ferromagnetic ground state}

The ferromagnetic ground states of $\hat{ H}$ is given by $|F\rangle = |\downarrow\downarrow\ldots\downarrow\downarrow\rangle$ such that all spins are pointing in the $\hat{z}$-direction. The energy of the ferromagnetic ground state is 
\be
\hat{ H}|F\rangle = E_F |F\rangle, \quad E_F = -2NJS^2 - N\Delta.
\ee
Furthermore, the ground state satisfies ${\hat S}_{j}^- |F\rangle = 0$, and ${\hat S}_j^z |F\rangle = -S|F\rangle$. In momentum space, these two relations become 
\be
\hat{S}_{\bm k}^-|F\rangle = 0, \quad \hat{S}_{\bm k}^z|F\rangle = -S\sqrt{N}\delta_{\bm k,0}|F\rangle. 
\label{seq:Fproperties}
\ee

\subsection{One magnon eigenstates}

There is a total of $N$ possible ways to do a single spin flip over the ferromagnetic ground state, $S_i^+|F\rangle$ for $i = 1,\ldots,N$, giving rise to a total of $N$ one-magnon eigenstates. Single magnon eigenstates of $\hat{ H}$ are {\it exactly} given by $|{\bm k}\rangle = \hat{S}_{\bm k}^+|F\rangle$. To show that this is the case, we note that $\hat{ H}\hat{S}_{\bm k}^+|F\rangle = [\hat{S}_{\bm k}^+\hat{ H} + \hat{R}_{\bm k}]|F\rangle$, where 
\be
\hat{R}_{\bm p} = [ \hat{ H},{\hat S}_{\bm p}^+] = \frac{J}{\sqrt{N}} \sum_{\bm q} (\gamma_{\bm p} - \gamma_{{\bm p} - {\bm q}})\left[ {\hat S}_{\bm q}^z{\hat S}_{{\bm p} - {\bm q}}^+ - {\hat S}_{\bm q}^+{\hat S}_{{\bm p} - {\bm q}}^z\right] + \Delta S_{\bm p}^+. 
\ee
Using Eq.(\ref{seq:Fproperties}), we find $\hat{R}_{\bm k}|F\rangle = [\Delta + JS(\gamma_0 - \gamma_{\bm k})] \hat{S}_{\bm k}^+|F\rangle$. As a result, 
\be
\hat{ H}\hat{S}_{\bm k}^+|F\rangle = [E_F + \varepsilon_{\bm k}]\hat{S}_{\bm k}^+|F\rangle,\quad \varepsilon_{\bm k} = \Delta + JS(\gamma_0-\gamma_{\bm k}),
\label{seq:kinetic}
\ee
and $|{\bm k}\rangle = \hat{S}_{\bm k}^+|F\rangle$ is an eigenstate of the Hamiltonian with energy $\varepsilon_{\bm k} $ over the vacuum energy. Because $\langle {\bm k}|{\bm p}\rangle = \delta_{\bm k,\bm p}$, the one magnon eigenstates $|\bm k\rangle = \hat{S}_{\bm k}^+|F\rangle$ are already normalized. 

\subsection{Two magnon eigenstates}

Spin wave theory assumes that $M$ magnon eigenstates are superposition of $M$ one-magnon eigenstates, for instance 
\be
|{\bm k},{\bm p}\rangle = \frac{1}{\sqrt{2SM_{\bm k}}\sqrt{2SM_{\bm p}}}\hat{S}_{\bm k}^+\hat{S}_{\bm p}^+|F\rangle, 
\label{seq:2magnons}
\ee
for $M_{\bm k}+M_{\bm p} = 2$. Such a basis has several problems, even in the simplest case $M=2$. First, the two-magnon basis in Eq.(\ref{seq:2magnons}) is overcomplete for $S=1/2$. In particular, for $S=1/2$, there is a total of $N(N-1)/2$ ways to do two spin flips on the lattice, giving rise to $N(N-1)/2$ two-magnon eigenstates of $\hat{ H}$. However, there are in total $N(N+1)/2$ ways in which ${\bm k},{\bm p}$ pairs of momenta can be chosen (such problem is not present for $S>1/2$).

Secondly, the two-magnon basis in Eq.(\ref{seq:2magnons}) is neither orthogonal nor properly normalized. Indeed, the scalar product of two elements of the basis is given by
\be
\langle F | \hat{S}_{-\bm p'}^-\hat{S}_{-\bm k'}^-\hat{S}_{\bm k}^+\hat{S}_{\bm p}^+|F\rangle = (2S)^2(\delta_{\bm k,\bm k'}\delta_{\bm p,\bm p'} + \delta_{\bm k,\bm p'}\delta_{\bm p,\bm k'} - \delta_{\bm k+\bm p, \bm k' + \bm p'}/N),
\label{seq:normalization1}
\ee
for $S=1/2$, and
\be
\langle F | \hat{S}_{-\bm p'}^-\hat{S}_{-\bm k'}^-\hat{S}_{\bm k}^+\hat{S}_{\bm p}^+|F\rangle = (2S)^2(\delta_{\bm k,\bm k'}\delta_{\bm p,\bm p'} + \delta_{\bm k,\bm p'}\delta_{\bm p,\bm k'}) + 4S(S-1)\delta_{\bm k+\bm p, \bm k' + \bm p'}/N,
\label{seq:normalization2}
\ee
for $S>1/2$. As a result, orthogonality and normalization of the two magnon basis (\ref{seq:2magnons}) is valid up to terms ${ O}(1/N)$. 

Finally, and crucial for our discussion, the two-magnon basis (\ref{seq:2magnons}) are not an eigenstates of the Heisenberg Hamiltonian. In particular, the effect of acting $\hat{ H}$ on a two magnon state ${\hat S}_{\bm p}^+{\hat S}_{\bm k}^+|F\rangle$ is given by
\be
\hat { H} {\hat S}_{\bm p}^+{\hat S}_{\bm k}^+|F\rangle = \left[{\hat S}_{\bm p}^+{\hat S}_{\bm k}^+{ H} + {\hat S}_{\bm p}^+{\hat R}_{\bm k} + {\hat S}_{\bm k}^+{\hat R}_{\bm p}\right]|F\rangle + {\hat Q}_{{\bm p}{\bm k}}|F\rangle , 
\ee
where we defined 
\be
{\hat Q}_{{\bm p}{\bm k}} = \left[\left[ \hat{ H}_J , {\hat S}_{\bm p}^+ \right],{\hat S}_{\bm k}^+\right] = \frac{J}{N}\sum_{\bm q} \left(\gamma_{\bm q}-\gamma_{{\bm q}-{\bm p}}-\gamma_{{\bm q}+{\bm k}}+\gamma_{{\bm q}-{\bm p}+{\bm k}}\right){\hat S}_{{\bm k}+{\bm q}}^+{\hat S}_{{\bm p}-{\bm q}}^+.
\label{seq:Rpk}
\ee
Using Eq.(\ref{seq:Fproperties}), we find that 
\be
\hat{ H} {\hat S}_{\bm p}^+{\hat S}_{\bm k}^+|F\rangle = \left[E_F + JS(\gamma_0-\gamma_{\bm k})+JS(\gamma_0-\gamma_{\bm p})\right]{\hat S}_{\bm p}^+{\hat S}_{\bm k}^+|F\rangle + {\hat Q}_{{\bm p}{\bm k}}|F\rangle,
\ee
where the first term on the right-hand side is the usual spin wave contribution which is diagonal on the two-magnon basis in Eq.(\ref{seq:2magnons}). The second term (${\hat Q}_{{\bm p}{\bm k}}$), however, creates two magnon states with momenta ${\bm p}+{\bm q}$ and ${\bm k} - {\bm q}$, for all ${\bm q}$. As a result, different two-magnon states are coupled by the matrix elements 
\be
\langle F | \hat{S}_{\bm k + \bm q}^-\hat{S}_{\bm p - \bm q}^-\hat{ H}\hat{S}_{\bm k }^+\hat{S}_{\bm p}^+|F \rangle = \frac{J}{N}\left(\gamma_{\bm q}-\gamma_{{\bm q}-{\bm p}}-\gamma_{{\bm q}+{\bm k}}+\gamma_{{\bm q}-{\bm p}+{\bm k}}\right) \approx \frac{Ja^2}{N} ({\bm k}\cdot{\bm p}),
\ee
where we evaluated the matrix elements at low momenta. 

Rather than dealing with the complications introduced by the two-magnon basis described above, it is possible to calculate exactly the two-magnon eigenstate and compute the matrix element in terms of single magnon eigenstates $|\bm k\rangle$. This provides a minimal description of the interacting magnon fluid at small densities, $n a^2 \ll 1$. 
The wavefunction for two-magnon states can be generically written as
\be
|\psi\rangle = \sum_{ij} \psi_{i,j}{\hat S}_i^+{\hat S}_j^+| F \rangle,
\ee
where $\psi_{i,j} = \psi_{j,i}$. In the case $S=1/2$, we do not need to explicitly set $\psi_{i,i} = 0$ because ${\hat S}_i^+{\hat S}_i^+|F\rangle $ already gives 0. To find the eigenstates of $\hat{ H}$, we act $\hat{ H}$ on two spin operators, 
\be
\hat{ H}{\hat S}_i^+{\hat S}_j^+ |F\rangle = \left[E_F{\hat S}_i^+{\hat S}_j^+ + 2JS {\hat S}_i^+\sum_\tau({\hat S}_{j+\tau}^+ - {\hat S}_j^+) + 2JS {\hat S}_j^+\sum_\tau({\hat S}_{i+\tau}^+ - {\hat S}_i^+) + 2J{\hat S}_i^+{\hat S}_j^+ \delta_{\langle i,j \rangle} - \delta_{ij} {\hat S}_i^+\sum_\tau {\hat S}_{i+\tau}^+\right]|F\rangle.
\ee
Taking the scalar product with the vector $\langle F | {\hat S}_{l}^-{\hat S}_{m}^-$ gives rise to the eigenvalue equations
\be
2JS\sum_{\tau}(\psi_{l,m+\tau} - \psi_{l,m} + \psi_{l+\tau,m} - \psi_{l,m}) + \frac{J}{2}[\psi_{l,m}+\psi_{m,l}-\psi_{l,l}-\psi_{m,m}]\delta_{\langle l,m \rangle} = E \psi_{l,m}, 
\label{seq:eigenvalueequation}
\ee
valid for $m\neq l$, and 
\be
2JS (\psi_{l,l+1} - \psi_{l,l}) + 2JS(\psi_{l+1,l}-\psi_{l,l}) = E \psi_{l,l}, 
\ee
valid for $m = l$. In Eq.(\ref{seq:eigenvalueequation}), $\delta_{\langle l , m \rangle}$ is 1 if $l$ and $m$ are nearest neighbors and 0 otherwise, and $E$ is referenced from $E_F$. Because of periodic boundary conditions, we can write $\psi_{i,j}$ in the center of mass frame as $\psi_{i,j} = \frac{e^{i{\bm K}\cdot {\bm R}}}{N}\sum_{\bm q}\psi_{{\bm K},\bm q}e^{i {\bm q} \cdot {\bm r}}$, where ${\bm K} = {\bm k} + {\bm p}$, ${\bm q} = ({\bm k} - {\bm p})/2$, ${\bm R} = ({\bm r}_i + {\bm r}_j)/2$ and ${\bm r} = {\bm r}_i - {\bm r}_j$, which gives rise to the eigenvalue equations
\be
(\varepsilon_{\bm k} + \varepsilon_{\bm p} - E)\psi_{\bm K,\bm q} = \frac{J}{N}\sum_{\bm k \tau}\cos({\bm q}\cdot {\boldsymbol\tau})\left[\cos({\bm K}\cdot{\boldsymbol\tau}/2)-\cos({\bm k}\cdot{\boldsymbol\tau})\right]\psi_{\bm K,\bm k}.
\label{seq:eigenvalueeq}
\ee
This is the two-magnon eigenvalue equation in the center of mass frame. The eigenstates of Eq.(\ref{seq:eigenvalueeq}) can be found using the $S$-matrix approach. We first note that the exact eigenstates can be labeled with the momenta of the incoming magnons, $\bm k = {\bm K}/2 + {\bm q}_0$ and $\bm p = {\bm K}/2 - {\bm q}_0$. Under this picture, the matrix elements for two magnon scattering is given by 
\be
\langle F | \hat{S}_{\bm k + \bm q}^- \hat{S}_{\bm p - \bm q}^- | {\bm k},{\bm p}\rangle = \psi_{\bm K , \bm q},
\ee
and 0 for momentum non-conserving processes. 
Singling out ${\bm q}_0$ in Eq.(\ref{seq:eigenvalueeq}), the eigenstate equations for the remaining $\bm q$ vectors is given by 
\be
\lambda_{\bm q} \psi_{\bm q} = \frac{J}{N}\sum_{\bm p}\Gamma_{\bm q \bm p}\psi_{\bm q} + \frac{J}{N}\Gamma_{{\bm q} {\bm q}_0}, \quad \bm q \neq \pm {\bm q}_0,
\label{seq:eigenvalueeq2}
\ee
where we defined the quantities 
\be
\lambda_{\bm q} = \varepsilon_{\bm K/2+\bm q} + \varepsilon_{\bm K/2-\bm q}- E, \quad \Gamma_{\bm q \bm p} = \sum_{\boldsymbol\tau}\cos({\bm q}\cdot {\boldsymbol\tau})\left[\cos({\bm K}\cdot{\boldsymbol\tau}/2)-\cos({\bm k}\cdot{\boldsymbol\tau})\right],
\ee
and, for compactness, we removed the subindex ${\bm K}$ from all quantities. 
Equation (\ref{seq:eigenvalueeq2}) can be written more conveniently as
\be
\psi_{\bm q} = \frac{J}{N}\frac{1}{\lambda_{\bm q}}(\Lambda_{\bm q} + \Gamma_{\bm q{\bm q}_0}), 
\label{seq:solution0}
\ee
where $\Lambda_{\bm q}= \sum_{\bm p}\Gamma_{\bm q \bm p}\psi_{\bm p}$ satisfies the self-consistent equation
\be
\Lambda_{\bm q} = \frac{J}{N}\sum_{\bm p}\left( \Gamma_{\bm q \bm p}\frac{1}{\lambda_{\bm p}}\Lambda_{\bm p} + \Gamma_{\bm q \bm p}\frac{1}{\lambda_{\bm p}}\Gamma_{\bm p {\bm q}_0} \right), 
\ee
The exact solution for $\Lambda_{\bm k}$ is 
\be
\Lambda_{\bm q} = \sum_{\bm k \bm p}\left(1-\frac{J}{N}\Gamma_{\bm k\bm q}\frac{1}{\lambda_{\bm q}}\right)^{-1}\frac{J}{N}\Gamma_{\bm k \bm p}\frac{1}{\lambda_{\bm p}}\Gamma_{\bm p {\bm q}_0}. 
\ee
Using $\Lambda_{\bm q}$ into Eq.(\ref{seq:solution0}) results in the wavefunction in the center of mass frame:
\be
\psi_{\bm q} = \sum_{\bm p}\left( 1 - \frac{J}{N}\Gamma_{\bm p \bm q}\frac{1}{\lambda_{\bm q}}\right)^{-1} \frac{J}{N}\frac{1}{\lambda_{\bm q}}\Gamma_{\bm q {\bm q}_0}.
\ee
Within the Born approximation, the wavefunction can be approximated as $\psi_{\bm q} \approx \frac{J}{N}\frac{1}{\lambda_{\bm q}}\Gamma_{\bm q{\bm q}_0}$. Further, for small wavevectors of the incoming particles, we can approximate $\Gamma_{\bm q {\bm q}_0}\approx a^2({\bm k}\cdot{\bm p})$. As a result, the exact two magnon eigenstates (at low momenta of incoming particles) can be interpreted as the scattering states of two spin waves coupled by the bare interaction of the form $\langle F | \hat{S}_{\bm k + \bm q}^- \hat{S}_{\bm p - \bm q}^- | {\bm k},{\bm p}\rangle \approx Ja^2({\bm k}\cdot{\bm p})$. 

\begin{figure}
  \centering\includegraphics[scale=1.0]{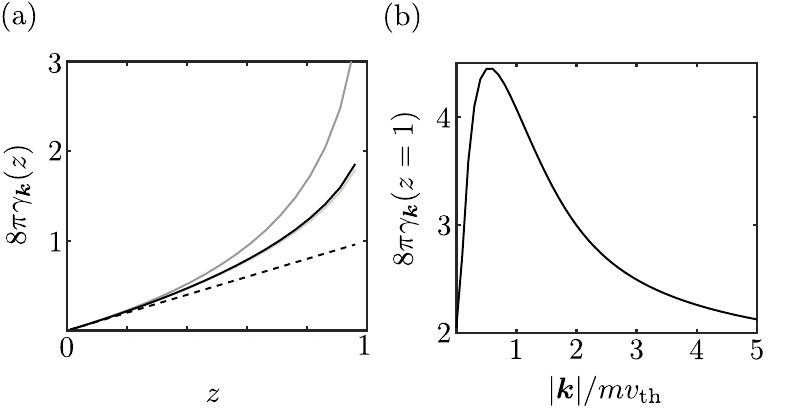}
  \caption{(a) $\gamma_{\bm k}(z)$ plotted for different values $|{\bm k}|/mv_{\rm th} = 0, 1, 5$ (increasing darkness). Indicated with dashed line is the linear $\gamma_{\bm k}(z) = z/8\pi$ obtained from the classical Boltzmann equation. (b) $\gamma_{\bm k}(z)$ exhibits a weak dependence on $k$, as shown for $z = 1$. At most, $\gamma_{\bm k}(z)$ varies by a factor of $\sim 2.5$ as $k$ is varied. In our calculations, we take the average of $\gamma_{\bm k}$ over $\bm k$ space.}
\label{fig:gamma}
\vspace{-4mm}
\end{figure}

\section{B. Relaxation time due to exchange coupling}

To estimate the relaxation time induced by the exchange interaction, we consider a magnon fluid at thermodynamic equilibrium and zero drift velocity, $\bar{n}_{\bm k} = 1/(z^{-1}e^{\varepsilon_{\bm k}/T}-1)$. Let us a assume that, at $t=0$, a non-equilibrium distribution is formed with a bump at wavevector ${\bm k}$, i.e. $n_{\bm p} = \bar{n}_{\bm p} + \delta n_{\bm k}\delta_{\bm k,\bm p}$. The relaxation time for such a distribution is given by 
\be
\frac{1}{\tau_{\bm k}} = \frac{(Ja^2)^2}{N^2} \sum_{{\bm p}{\bm q}} ({\bm k}\cdot{\bm p})^2 2\pi \delta(\varepsilon_{\bm k}+\varepsilon_{\bm p} - \varepsilon_{\bm k + \bm q} - \varepsilon_{\bm p - \bm q})\left[\bar{n}_{{\bm p}}(1+\bar{n}_{{\bm k}+{\bm q}})(1+\bar{n}_{{\bm p}-{\bm q}}) - (1 + \bar{n}_{{\bm p}})\bar{n}_{{\bm k}+{\bm q}}\bar{n}_{{\bm p}-{\bm q}}\right].
\label{sec:relaxationtime}
\ee
The relaxation time can be expressed as $\frac{1}{\tau_{\bm k}} = \frac{\gamma_{\bm k}(z)}{16\pi}\frac{T^2(ka)^2}{J}$ after factoring out the ${\bm k}$ vector dependence out of the integral, normalizing energies with $T$, and momenta with $\sqrt{2mT}$. The dimensionless prefactor $\gamma_{\bm k}(z)$ is plotted in Fig.\ref{fig:gamma}, exhibits a weak dependence on $\bm k$, and scales approximately as $\propto z$. Rather than keeping this unimportant ${\bm k}$ dependence of ${\gamma}_{\bm k}$, we define an average ${\gamma}$ of all ${\bm k}$ vectors and $z$ values, ${\gamma}(z)/z = \int_0^1 dz/z \int d^2\tilde{\bm k}/(2\pi)^2 {\gamma}_{\bm k}(z)$, which yields $\gamma(z)\approx cz$, with $c\sim {\cal O}(1)$.

In thermal equilibrium, the typical relaxation rate for thermal magnons is given by $1/\bar{\tau} \sim \frac{T^2(na^2)}{J}$. The relaxation length of thermal magnons is given by $\ell = \bar{v}\bar{\tau}$, where $\bar{v}^2 = \frac{1}{2\pi mn}\int_0^\infty d{k}{k}^3\bar{n}_{\bm k}=2mTg_2(z)/g_1(z)$ is the thermal velocity, and results in Eq.(\ref{eq:relaxationlength}) of the main text.

\section{C. Derivation of hydrodynamic equations}

In this section we derive the current operators associated with the conserved quantites of the effective Hamiltonian 
\be
\hat{H} = \int_{\bm x} \frac{\partial_\alpha\hat{\psi}_{\bm x}^\dagger \partial_\alpha\hat{\psi}_{\bm x}}{2m}+g\left(\hat{\psi}_{\bm x}^\dagger\hat{\psi}_{\bm x}^\dagger\partial_\alpha\hat{\psi}_{\bm x}\partial_\alpha\hat{\psi}_{\bm x}+{\it h.c.}\right),
\label{seq:Heffective}
\ee
which was derived in the main text. Here we defined $g = Ja^2/4$ for compactness of notation. We recall that the Hamiltonian (\ref{seq:Heffective}) has three conserved quantities: particle number $\hat{N} = \int_{\bm x} \hat{n}_{\bm x} = \int_{\bm x} \hat{\psi}_{\bm x}^\dagger\hat{\psi}_{\bm x}$, momentum $\hat{P}_{\alpha} = \int_{\bm x}\hat{p}_{\alpha,\bm x} = \frac{-i}{2}\int_{\bm x}\hat{\psi}_{\bm x}^\dagger \partial_\alpha\hat{\psi}_{\bm x}-{h.c.}$, and energy $\hat{ H} = \int_{\bm x}\hat{\epsilon}_{\bm x}$. We proceed to derive the currents associated with each of the conserved quantities. 

\subsection{Current operators}

The current operators can be derived from the continuity relation that ensures charge conservation: $\partial_t\hat{n}_{\bm x} = - \partial_\alpha \hat{J}_\alpha = i[\hat{H},\hat{n}_{\bm x}]$ for particle number, $\partial_t \hat{p}_{\alpha,\bm x} = -\partial_\alpha \hat{\Pi}_{\alpha,\beta} = i[\hat{H},p_{\beta,\bm x}]$ for momentum, and $\partial_t \hat{\epsilon}_{\bm x} = -\partial_\alpha \hat{Q}_\alpha = i [\hat{H},\hat{\epsilon}_{\bm x}]$ for energy. Computing the commutator of $\hat{H}$ with each of the local operators gives rise to the currents:
\be
\begin{array}{rl}
   \hat{J}_{\alpha} = & \displaystyle\frac{-i}{2m}\left[\hat\psi_{\bm x}^\dagger\partial_\alpha\hat\psi_{\bm x}-h.c.\right] + 2ig\left[ \hat\psi_{\bm x}^\dagger \hat\psi_{\bm x}^\dagger \partial_\alpha\hat\psi_{\bm x}\hat\psi_{\bm x} -h.c. \right],\\ & \\
   \hat{\Pi}_{\alpha\beta} = & \displaystyle\frac{1}{2m}\left[ \partial_\alpha\hat{\psi}_{\bm x}^\dagger\partial_\beta\hat{\psi}_{\bm x} +h.c.\right] + g \left[(\hat{\psi}_{\bm x}\partial_{\gamma}\hat{\psi}_{\bm x})^2\delta_{\alpha\beta}+2\hat{\psi}_{\bm x}^\dagger\hat{\psi}_{\bm x}^\dagger\partial_\alpha\hat{\psi}_{\bm x}\partial_\beta\hat{\psi}_{\bm x}+h.c.\right],\\ & \\
   \hat{Q}_\alpha = &\displaystyle\frac{-i}{4m}\left[\partial_\beta\hat{\psi}_{\bm x}^\dagger\partial_\alpha\partial_\beta\hat{\psi}_{\bm x} -h.c.\right] - \frac{ig}{m}\left[\hat{\psi}_{\bm x}^\dagger\hat{\psi}_{\bm x}^\dagger\partial_\beta\hat{\psi}_{\bm x}\partial_\alpha\partial_\beta\hat{\psi}_{\bm x} + \hat{\psi}_{\bm x}^\dagger\hat{\psi}_{\bm x}^\dagger\partial_\alpha\hat{\psi}_{\bm x}\partial_\beta^2\hat{\psi}_{\bm x} - \partial_\alpha\hat{\psi}_{\bm x}^\dagger\hat{\psi}_{\bm x}^\dagger\partial_\beta\hat{\psi}_{\bm x}\partial_\beta\hat{\psi}_{\bm x}-h.c.\right].
\end{array}
\ee

\subsection{Currents within the gaussian approximation}

We compute the expectation value of the currents using the {\it local equilibrium} approximation which allows us to coarse-grain real space in regions of size $\ell$ in which the system is effectively thermalized. We also employ the gaussian approximation to represent the density matrix in the subregion region $\bm x$ as $\hat\rho_{\bm x} = {\rm exp}\left(-\sum_{\bm k}\frac{\bm k^2}{2mT}\psi_{\bm k}^\dagger\psi_{\bm k} - u_{\alpha}\hat{P}_\alpha-\mu\hat{N}\right)$, where we use the bare mass $m$ rather than the renormalized mass for simplicity. The gaussian approximation enables us to compute four-point correlation functions in terms of two-point correlations function. In particular, the expectation value of the currents is given by 
\be
\begin{array}{cc}
  \displaystyle  n = \langle 1 \rangle & \displaystyle J_\alpha = \frac{\langle k_\alpha \rangle}{m} - 8g\langle 1 \rangle\langle k_\alpha\rangle, \\ & \\ 
  \displaystyle P_\alpha = \langle k_\alpha\rangle , &\displaystyle \Pi_{\alpha\beta} = \frac{\langle k_\alpha k_\beta\rangle}{m} - 4g\langle k_\gamma\rangle\langle k_\gamma\rangle \delta_{\alpha\beta} - 8g \langle k_\alpha\rangle\langle k_\beta\rangle, \\ & \\
  \displaystyle \epsilon = \frac{\langle k_\beta k_\beta\rangle}{2m} - 4g \langle k_\beta\rangle\langle k_\beta\rangle, & \displaystyle \quad Q_\alpha = \frac{\langle k_\alpha k_\beta k_\beta\rangle }{2m^2} - 8g \langle k_\beta \rangle \frac{\langle k_\alpha k_\beta\rangle }{m} - 4g\langle k_\alpha\rangle \frac{\langle k_\beta k_\beta\rangle}{m},
  \label{seq:expcurrents}
\end{array}
\ee
where we used the short-hand notations $\langle A \rangle = \int \frac{d{\bm k}}{(2\pi)^2}A_{\bm k}n_{\bm k}$, and $n_{\bm k}$ is the Bose distribution function with chemical potential $\mu$, drift velocity $u_\alpha$, and temperature $T$. It is straight-forward to compute the expectation values, which are given by: $\langle 1 \rangle = mTg_1(z)/2\pi$, $\langle k_\alpha \rangle = nmu_\alpha$, $\langle k_\alpha k_\beta \rangle = mnu_\alpha u_\beta + \langle \tilde{k}_\alpha\tilde{k}_\beta\rangle$, $\langle k_\alpha k_\beta k_\beta \rangle = \langle \tilde{k}_\alpha \tilde{k}_\beta \tilde{k}_\beta \rangle + mu_\alpha\langle \tilde{k}_\beta \tilde{k}_\beta\rangle + 2mu_\beta \langle\tilde{k}_\alpha\tilde{k}_\beta\rangle + m^3 n u^2 u_\alpha$ (here $g_\nu(z)$ is the bose integral defined in the main text, and $\tilde{k}_{\alpha} = k_\alpha - mu_\alpha$). The term $\langle \tilde{k}_\alpha \tilde{k}_{\beta}\rangle = P_{\alpha\beta} = n\theta\delta_{\alpha\beta}+P_{\alpha\beta}'$ is the pressure tensor with $P_{\alpha\beta}'$ the dissipative component, and $\langle \tilde{k}_{\alpha}\tilde{k}_{\beta}\tilde{k}_{\beta}\rangle = q_\alpha$ is the heat current. Both $P_{\alpha\beta}'$ and $q_\alpha$ are estimated below. Replacing the expectation values into Eq.(\ref{seq:expcurrents}) results in the charges and currents:
\be
\begin{array}{cc}
  \displaystyle n, &  J_\alpha = n v_\alpha, \\ & \\
  \displaystyle p_\alpha = mu_\alpha, & \Pi_{\alpha\beta} = P_{\alpha\beta}+np_\alpha v_\beta,\\ & \\
\displaystyle  \epsilon = \frac{d\theta}{2} +\frac{p_\alpha v_\beta}{2}, & \quad Q_\alpha = q_\alpha + n\epsilon v_{\alpha} + P_{\alpha\beta}v_\beta. 
  \end{array}
\ee
The continuity equations $\partial n + \partial_\alpha J_\alpha = 0$, $\partial_t(np_\alpha) + \partial_\beta\Pi_{\alpha\beta} = 0$, and $\partial_t(n\epsilon) + \partial_\alpha Q_\alpha = 0 $ give rise to the hydrodynamic equations (\ref{eq:hydro}) in the main text. 

\section{D. Estimating transport coefficients from the relaxation time approximation} 

To compute the leading order corrections to $P_{\alpha\beta}$ and $q_\alpha$, we need to determine $\delta n_{\bm k}$ induced by gradients in $n$, $u_\alpha$, and $\theta$. With this objective in mind, we linearize the Boltzmann kinetic equation 
\be
\left( \partial_t + v_{\bm k,\alpha} \partial_\alpha + {F}_\alpha\partial_{k_\alpha} \right)\bar{n}_{\bm k} = { I}(\bar{n}_{\bm k}+\delta n_{\bm k}).
\label{seq:boltzmann1}
\ee
Here we assumed that $\delta n_{\bm k} \ll n_{\bm k}$, such that the leading order contributions on the left-hand is given by the derivatives (both space and time) of $\bar{n}_{\bm k}$. The right-hand side is already leading order in $\delta n_{\bm k}$ because ${ I}(\bar{n}_{\bm k}) = 0$. 

We begin the analysis by considering the left-hand side of Eq.(\ref{seq:boltzmann1}). We recall that $\bar{n}_{\bm k}(n,u_{\alpha},\theta)$ is the local distribution function which depends implicitly on $n$, $u_{\alpha}$ and $\theta$. As such, computing the time and spatial derivatives of $\bar{n}_{\bm k}$ leads to 
\be
\left[\partial_t + v_{\bm k,\alpha}\partial_\alpha\right] \bar{n}_{\bm k} = \left[ \dot{n} + v_{\bm k,\alpha} \partial_\alpha n\right] \partial_n \bar{n}_{\bm k}\big\rvert_{\theta,u_\alpha} + \left[ \dot{\theta} + v_{\bm k,\alpha} \partial_\alpha \theta \right]\partial_\theta \bar{n}_{\bm k} \big\rvert_{n,u_\alpha} +  \left[ \dot{u}_\alpha + v_{\bm k,\beta} \partial_\beta u_\alpha \right] \partial_{u_\alpha} \bar{n}_{\bm k}\big\rvert_{n,\theta}, 
\label{seq:lhs1}
\ee
where $\partial \bar{n}_{\bm k}/\partial x |_{y,z}$ denotes the derivative of $\bar{n}_{\bm k}$ with respect to $x$, leaving $y$ and $z$ constant. In Eq.(\ref{seq:lhs1}), we replace the time derivatives $\dot{n}$, $\dot{u}_{\alpha}$, and $\dot{\theta}$ by the hydrodynamic equations (\ref{eq:hydro}) of the main text in the local equilibrium approximation, and compute transport coefficients to leading order in $na^2$. {\it i.e.} using $P_{\alpha\beta} = \delta_{\alpha\beta}n\theta/m$ and $q_\alpha=0$. This results in 
\be
\begin{array}{r}
\displaystyle\left[\partial_t + v_{\bm k,\alpha}\partial_\alpha + F_{\alpha}\partial_{k_\alpha}\right] \bar{n}_{\bm k} = \left[\delta_{\alpha\beta}\partial_n \bar{n}_{\bm k}\big\rvert_{\theta,u_{\alpha}} + \frac{m}{n} \partial_nP_{\alpha\beta}\partial_{\theta_{\bm k}} \bar{n}_{\bm k}\right]\tilde{v}_{\bm k,\beta}\partial_\alpha n + \left[\delta_{\alpha\beta}\partial_\theta \bar{n}_{\bm k}\big\rvert_{n,u_{\alpha}} + \frac{m}{n}\partial_\theta P_{\alpha\beta}\partial_{\theta_{\bm k}} \bar{n}_{\bm k} \right]\tilde{v}_{\bm k,\beta}\partial_\alpha \theta  \\
\displaystyle - \left[\delta_{\alpha\beta} n \partial_n \bar{n}_{\bm k}\big\rvert_{\theta,u_\alpha} + \frac{m}{n}P_{\alpha\beta}\partial_\theta \bar{n}_{\bm k}\big\rvert_{n,u_{\alpha}} + m \tilde{v}_{\bm k,\alpha}\tilde{v}_{\bm k,\beta} \partial_{\theta_{\bm k}} \bar{n}_{\bm k}\right]\partial_\alpha u_{\beta},
\end{array} 
\label{seq:lhs2}
\ee
where we used the identities $\partial \bar{n}_{\bm k} / \partial u_\alpha  |_{n,\theta}= - [\partial \bar{n}_{\bm k}/\partial\theta_{\bm k}] m\tilde{v}_{\bm k,\alpha} $ and $F_\alpha\partial_{k_\alpha}{\bar n}_{\bm k} = F_\alpha[\partial {\bar n}_{\bm k}/\partial{\theta_{\bm k}}]\tilde{v}_{\bm k,\alpha}$. The terms in brackets in Eq.(\ref{seq:lhs2}) are thermodynamic functions that depend on the local values of $(T,z,w_{\alpha})$ and are given by 
\be
\begin{array}{r}
\displaystyle\left[\partial_t + v_{\bm k,\alpha}\partial_\alpha + F_\alpha\partial_{k_\alpha}\right] \bar{n}_{\bm k} = \left[ \frac{2\pi}{mT}\left(h_n(z) + \tilde{h}_n(z) \frac{\theta_{\bm k}}{T}\right)\tilde{v}_{{\bm k},\alpha}\partial_\alpha n + \frac{2\pi}{T}\left(h_\theta(z) + \tilde{h}_\theta(z) \frac{\theta_{\bm k}}{T}\right)\tilde{v}_{{\bm k},\alpha}\partial_\alpha \theta \right. \\ \\
\displaystyle\left. + \left(\delta_{\alpha\beta}\frac{\theta_{\bm k}}{T} - \frac{mv_{\bm k,\alpha} v_{\bm k,\beta}}{T}\right) \partial_\alpha u_\beta \right] \bar{n}_{\bm k}(\bar{n}_{\bm k}+1),
\end{array} 
\ee
where the dimensionless coefficients $h_{n,\theta}(z)$ and $\tilde{h}_{n,\theta}(z)$ are
\be
\begin{array}{l}
\displaystyle h_n(z) = \frac{zg_2^2 - (1-z)g_2g_1^2}{zg_2g_1^2 - (1-z)g_1^4/2}, \quad  \tilde{h}_n(z) = \left[\frac{1}{g_1}+\frac{zg_2}{g_1^2(1-z) - 2zg_2}\right],\\ \\
\displaystyle h_\theta(z) = \frac{zg_2^2 - (1-z)g_2g_1^2}{zg_2g_1^2 - (1-z)g_1^4/2}, \quad  \tilde{h}_\theta(z) = \left[\frac{1}{g_1}+\frac{zg_2}{g_1^2(1-z) - 2zg_2}\right].
\end{array}
\ee

Let us now focus on the right-hand side of Eq.(\ref{seq:boltzmann1}). There are many schemes to calculate ${ I}[\bar{n}_{\bm k} + \delta n_{\bm k}]$. The simplest approach is to use the {\it relaxation time approximation}. In this approximation, the collision integral is written as ${ I}[\bar{n}_{\bm k} + \delta n_{\bm k}] \approx - \delta n_{\bm k}/\tau_{\bm k}$, where $\tau_{\bm k}$ is defined in Eq.(\ref{sec:relaxationtime}). Importantly, we keep the explicit dependence on magnon wavevector. We note that $1/\tau_{\bm k}$ was calculated using $u_{\alpha} = 0$, but its value remains valid so long as $u_{\alpha} \lesssim \sqrt{T/m}$ [corrections to $1/\tau_{\bm k}$ due to finite drift velocity are ${ O}(u_{\alpha}^2)$]. As a result, $\delta n_{\bm k}$ becomes proportional to gradients in $n$, $\theta$, and $u_\alpha$:
\be
\begin{array}{rl}
\displaystyle\delta n_{\bm k} = & \displaystyle\tau_{\bm k} \left[ \frac{2\pi}{mT}\left(h_n(z) + \tilde{h}_n(z) \frac{\theta_{\bm k}}{T}\right)\tilde{v}_{{\bm k},\alpha}\partial_\alpha n + \frac{2\pi}{T}\left(h_\theta(z) + \tilde{h}_\theta(z) \frac{\theta_{\bm k}}{T}\right)\tilde{v}_{{\bm k},\alpha}\partial_\alpha \theta \right. \\ \\
& \displaystyle\left. + \left(\delta_{\alpha\beta}\frac{\theta_{\bm k}}{T} - \frac{mv_{\bm k,\alpha} v_{\bm k,\beta}}{T}\right) \partial_\alpha u_\beta \right] \bar{n}_{\bm k}(\bar{n}_{\bm k}+1).
\end{array}
\ee
Using $n_{\bm k} = \bar{n}_{\bm k} + \delta n_{\bm k}$ 
and integrating over ${\bm k}$ leads to 
\be
P_{\alpha\beta} = \frac{n\theta}{m}\delta_{\alpha\beta} + \mu\left(\partial_\alpha u_\beta + \partial_\beta u_\alpha \right) - \mu \delta_{\alpha\beta}\partial_\gamma u_\gamma, \quad q_\alpha = \kappa_n \partial_\alpha n + \kappa_\theta \partial_{\alpha}\theta,
\label{seq:pq}
\ee
where only linear terms on $\partial_\alpha n$, $\partial_\alpha\theta$, and $\partial_\alpha u_\beta$ were considered (i.e., gradients of thermodynamic quantities are small). For a two-dimensional magnon gas with quadratic dispersion and collision rate of the form $1/\tau_{\bm k}\propto {\bm k}^2$ (i.e., only considering exchange coupling), the relaxation time yields 
that dissipation is dominated by viscosity $\mu (T,z) \sim \frac{J^2}{T}$, while $\kappa_{n} = \kappa_{\theta} $ are second order effects (in powers of $T/J$) compared to $\mu$. $\kappa_n $ and $\kappa_\theta$ are dominated by deviations to quadratic dispersion and/or finite scattering at low scattering, e.g. dipolar interactions. 

\section{E. Transverse spin fluctuations}

The spectral weight of the correlator $\chi_{+-}({\bm k,\omega}) = -i \int_0^\infty dt e^{i\omega t} \langle [\hat{S}_{-\bm k}^-,S_{\bm k}^+(0)]\rangle$ is concentrated at the magnon frequency $\omega_{\bm k} = \Delta + \varepsilon_{\bm k}$ and is associated to the production of a single magnon. Off-resonant processes, however, give rise to a finite contribution to $\chi_{-+}({\bm k},\omega)$ below the magnon gap, see Fig.\ref{fig:diagram}(a). As such, we estimate the contribution of such processes in the noise spectrum and show that they give a small contribution to $\chi_{+-}$ compare to that of the sound mode. With this objective in mind, we calculate the leading order contribution of the imaginary part of the magnon self-energy $\Sigma ({\bm k},\omega)$, and approximate the correlation function as 
\be
\chi_{+-}({\bm k},\omega) = \frac{1}{\omega-\omega_{\bm k}+i\Sigma''({\bm k},\omega)}, 
\ee
where energy shifts to the single magnon dispersion are neglected. From the effective interaction in Eq.(\ref{eq:Heffective}) of the main text, this is given by the second order process depicted in Fig.\ref{fig:diagram}(b). In terms of Matsubara frequencies, it can be written as
\be
\Sigma({\bm k},\omega) = -J^2a^4\sum_{\bm p \bm q}\sum_{i\omega_n'i\omega_n''}({\bm k}\cdot{\bm p})^2 \frac{1}{(i\omega_n'-\omega_{\bm p})(i\omega_n+i\omega_n''-\omega_{\bm k + \bm q})(i\omega_n' -i\omega_n'' - \omega_{\bm p - \bm q})}. 
\ee
The retarded correlator is obtained by analytical continuatio $i\omega_n \rightarrow \omega + i\epsilon$ and taking the imaginary part of the resulting expression:
\be
\Sigma''(\bm k,\omega) = J^2a^4 \sum_{\bm p \bm q}({\bm k}\cdot{\bm p})^2 \delta(\omega -\Delta + \varepsilon_{\bm p} - \varepsilon_{\bm k + \bm q} - \varepsilon_{\bm p - \bm q})(n_{\bm p}-\tilde{n}_{\bm p})(1+n_{\bm k + \bm q}+n_{\bm p - \bm q}),
\label{seq:sigma}
\ee
where we denote $\tilde{n}_{\bm p} = n(\varepsilon_{\bm p} + \omega)$. A similar analysis follows for the correlator $\chi_{+-}(\omega) \approx \delta(\omega+\omega_{\bm k})$. Dimensional analysis in the limit $\omega \ll T$ yields $\Sigma''$ scaling as $\Sigma({\bm q},\omega) = \frac{T\omega}{J}(qa)^2$. 

\begin{figure}
  \centering\includegraphics[scale=1.0]{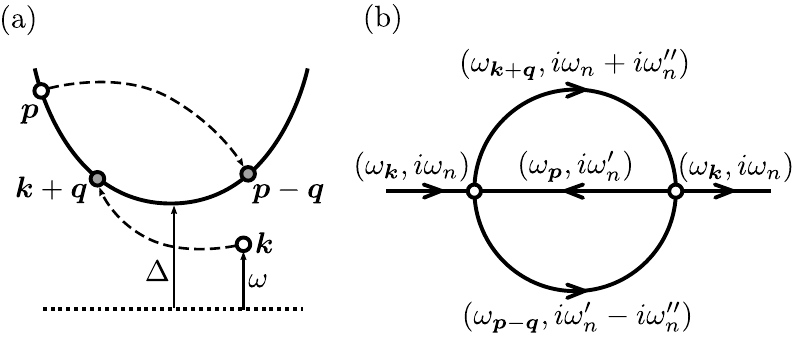}
  \caption{(a) In addition to the sound mode, off-resonant processes can also give a finite contribution to $\chi_{-+}$ below the magnon gap. (b) Sunrise diagram contributing to the magnon self-energy of $\chi_{-+}$.}
\label{fig:diagram}
\vspace{-4mm}
\end{figure} 

\section{F. Measurement of magnon sound modes}

We consider a spin-1/2 qubit with an intrinsic level splitting $\omega$ placed a distance $d$ above the magnetic insulator. The dynamics of the qubit and the ferromagnet is governed by the Hamiltonian $\hat{ H}_{\rm total} = \hat{ H} + \hat{ H}_{\rm c} + \hat{ H}_{\rm q}$. Here ${ H}_{\rm F}$ is the Hamiltonian of the ferromagnet, see main text. The term $\hat{ H}_{\rm q}$ is the qubit Hamiltonian given by $\hat{ H}_{\rm q} = \frac{1}{2}\omega{\bm n}_{\rm q}\cdot{\boldsymbol\sigma}$, where ${\bm n}_{\rm q}$ is the intrinsic polarizing field of the spin probe. For instance, in the case of NV centers in diamond, ${\bm n}_{\rm q}$ is the axis of the NV defect in the diamond lattice. Finally, the term $\hat{ H}_{\rm c}$ is the qubit-ferromagnet coupling, given by 
\be
{ H}_{\rm c} = \frac{\mu_{\rm B}^2}{2}\hat{\boldsymbol\sigma}\cdot\hat{\bm B}, \quad \hat{\bm B} =\frac{1}{4\pi} \sum_j \left[\frac{\hat{\bm S}_j}{r_j^3} - \frac{3(\hat{\bm S}_j\cdot{\bm r}_j) {\bm r}_j}{r_j^5}\right],
\label{seq:HqF}
\ee
where ${\bm B}$ is the magnetic field at the position of the probe induced by dipolar interactions with the 2D ferromagnet, and ${\bm r}_j = (x_j,y_j,-d)$ is the relative position between the $i$-th spin in the 2D lattice and probe. 

In thermal equilibrium, the 2D ferromagnet is described by the density matrix $\rho_{\rm F} = \sum_n e^{-\varepsilon_n/k_{\rm B}T}|n\rangle\langle n|$, where $|n\rangle$ are the eigenstates of the ferromagnet. The absorption rate, $1/T_{1,{\rm abs}}$, and emission rate, $1/T_{1,{\rm em}}$, is obtained from Fermi Golden's rule using the initial state $|i\rangle = |-\rangle\otimes\rho_{\rm F}$ and $|i\rangle = |+\rangle\otimes\rho_{\rm F}$, respectively: 
\be
1/T_{\rm abs,em} = 2\pi \sum_{nm} \rho_n {B}_{nm}^\pm{B}_{mn}^\mp \delta (\omega \pm \varepsilon_{mn}). 
\ee
Here ${B}_{nm}^\alpha$ denotes $\langle n | {\hat B}^\alpha | m \rangle$, and $\varepsilon_{mn}$ is the energy difference between states $m$ and $n$, $\varepsilon_{mn} = \varepsilon_m - \varepsilon_n$. The relaxation rate is defined as $1/T_{1} = \frac{1}{2}[1/T_{\rm abs} + T_{\rm em}]$. More compactly, $1/T_{1}$ can be expressed as
\be
\frac{1}{T_1} = \frac{\mu_{\rm B}^2}{2}\int_{-\infty}^{\infty} dt e^{i\omega t}\langle \{ \hat{B}^-(t), \hat{B}^+(0)\}\rangle.
\label{seq:T1}
\ee
For computation it is more convenient to express $1/T_1$ in terms of retarded correlation functions. In this direction, the fluctuation-dissipation theorem reads 
\be
\int_{-\infty}^{\infty} dt e^{i\omega t}\langle\{ \hat{B}^-(t),\hat{B}^+(0)\}\rangle = {\rm coth}\left(\frac{\omega}{2T}\right){\rm Im}\left[\chi_{B^-B^+}^{\rm R}(\omega)\right],
\label{seq:fl-dis}
\ee
where $\chi_{B^-B^+}^{\rm R}(\omega) = -i\int_0^\infty dt \langle[\hat{B}^-(t),\hat{B}^+(0)]\rangle$ is the retarded correlation function. 

Finally, $1/T_1$ can be expressed in terms of spin-spin correlation functions. Expressing $\hat{S}_{\tau}^\alpha = \sum_{\bm k}\frac{e^{i{\bm k}\cdot{\boldsymbol\tau}}}{\sqrt{N}}\hat{S}_{\bm k}^\alpha$ in momentum space and inserting into Eq.(\ref{seq:HqF}), we can express $\hat{B}^\alpha $ in terms of $S_{\bm k}^\pm$ and $S_{\bm k}^z$. Without loss of generality, we assume ${\bm k} = (k,0)$. For $\hat{B}^x$, we find
\be
\hat{B}_{\bm k}^x = \sum_{j}e^{ikx_j}\left[\left(\frac{1}{r_j^3}-\frac{3x_j^2}{r_j^5}\right)S_{\bm k}^x-\frac{3x_jy_j}{r_j^5}S_{\bm k}^y+\frac{3x_jd}{r_j^5}S_{\bm k}^z\right].
\label{seq:BxS}
\ee 
Using the continuum approximation to approximate $\sum_{j} \rightarrow \frac{1}{a^2}\int d^2{\bm x}$, the first term on the right-hand side of Eq.(\ref{seq:BxS}) is
\be
\sum_{j}e^{ikx_j}\left(\frac{1}{r_j^3}-\frac{3x_j^2}{r_j^5}\right) \rightarrow \frac{1}{a^2}\iint dxdy \,e^{ikx}\left(\frac{1}{r^3}-\frac{3x^2}{r^5}\right) = \frac{2}{a^2}\int dx e^{ik x}\frac{d^2 - x^2}{(d^2+x^2)^2} = \frac{2}{da^2}\int d\xi e^{i(kd)\xi}\frac{1-\xi^2}{(1+\xi^2)^2}.
\ee
In the last step, we can use the residue theorem to express $\int_{-\infty}^{\infty} d\xi e^{i(kd)\xi}\frac{1-\xi^2}{(1+\xi^2)^2}$ as $\oint dz e^{i(kd)z}\frac{1-z^2}{(1+z^2)^2}= \pi(kd)e^{-kd}$, where for $kd>0$ we use a contour of integration in the upper-half complex plane. As a result, we obtain
\be
\sum_{j}e^{ikx_j} \left(\frac{1}{r_j^3}-\frac{3x_j^2}{r_j^5}\right) \approx \frac{ke^{-kd}}{2a^2},
\ee
exact in the continuum limit. For the second term on the right-hand side of Eq.(\ref{seq:BxS}), we find $\sum_{j}e^{ikx_j}\frac{x_jy_j}{r_j^5} = 0$. Finally, for the third term in the right-hand side of Eq.(\ref{seq:BxS}), we find
\be
3\sum_j e^{ikx_j} \frac{x_jd}{r_j^5} \approx \frac{3ikd}{a^2}\iint dx\,dy\frac{x^2}{r^5} =  \frac{ik}{2a^2}.
\ee

Repeating the same procedure for $\hat{B}^y$ and $\hat{B}^z$, and generalizing our results for a generic ${\bm k}=(k_x,k_y)$, we obtain $\hat{B}^\alpha = \frac{1}{\sqrt{N}}\sum_{\bm k}B_{\bm k}^\alpha$, with 
\be
\left(
\begin{array}{c}
{\hat B}_{\bm k}^x\\
{\hat B}_{\bm k}^y\\
{\hat B}_{\bm k}^z
\end{array}\right) = \frac{e^{-|\bm k|z}}{2a^2}\left(\begin{array}{ccc}
k_x^2/|{\bm k}| & k_xk_y/|{\bm k}| & ik_x \\
k_xk_y/|{\bm k}| & k_y^2/|{\bm k}| & ik_y \\
ik_x  & ik_y & -|{\bm k}| 
\end{array}\right)
\left(\begin{array}{c}
\hat{S}_{\bm k}^x \\ \hat{S}_{\bm k}^y \\ \hat{S}_{\bm k}^z
\end{array}\right).
\label{seq:BvsS}
\ee
The ${B}_{\bm k}^\pm = B_{\bm k}^x \pm i B_{\bm k}^y$ terms can be written as a function of ${S}_{\bm k}^\pm$ and $S_{\bm k}^z$ such that Eq.(\ref{seq:BvsS}) can be recasted as
\be
\left(
\begin{array}{c}
{\hat B}_{\bm k}^+\\
{\hat B}_{\bm k}^-\\
{\hat B}_{\bm k}^z
\end{array}\right) = \frac{e^{-|\bm k|z}}{2a^2}\left(\begin{array}{ccc}
 \\
|{\bm k}|/2 & (k_x+ik_y)^2/2|{\bm k}| & ik_x-k_y \\
(k_x-ik_y)^2/2|{\bm k}|  & |{\bm k}|/2 & ik_x+k_y \\
(ik_x+k_y)/2 & (ik_x-k_y)/2 & - |{\bm k}|
\end{array}\right)
\left(\begin{array}{c}
\hat{S}_{\bm k}^+ \\ \hat{S}_{\bm k}^- \\ \hat{S}_{\bm k}^z
\end{array}\right).
\label{seq:BxS2}
\ee
Using Eq.(\ref{seq:BxS2}) in Eq.(\ref{seq:T1}), the spin qubit relaxation time is given by
\be
\frac{1}{T_1} = {\rm coth}\left(\frac{\omega}{2T}\right) \frac{\mu_{\rm B}^2}{2a^2} \int \frac{d^2{\bm k}}{(2\pi)^2} e^{-2|{\bm k}|d} |{\bm k}|^2 \left[\chi_{-+}^{\rm R}({\bm k},\omega)+\chi_{+-}^{\rm R}({\bm k},\omega)+4\chi_{zz}^{\rm R}({\bm k},\omega)\right], 
\label{seq:relaxationtime}
\ee
where we denote $\chi_{\alpha\beta}^{\rm R}({\bm k},\omega) = -i\int_0^\infty dt \langle[\hat{S}_{-\bm k}^\alpha(t),\hat{S}_{\bm k}^\beta(0)]\rangle$. 

\section{G. Effect of dipolar interactions}

Dipolar interactions, which can be sizable in a two-dimensional ferromagnet, introduce a variety of effects that need to be carefully taken into account, namely, it modifies the collision term by adding hard-core repulsion, and induce magnon leakage via three body interactions. We incorporate dipolar interactions via the term
\be
\hat{ H}_{\rm d} = \frac{\mu_{\rm B}^2}{4\pi}\frac{1}{2} \sum_{jj'}\left[\frac{\hat{\bm S}_j\cdot\hat{\bm S}_{j'}}{r_{jj'}^3}- 3 \frac{(\hat{\bm S}_j\cdot{\bm r}_{jj'})(\hat{\bm S}_{j'}\cdot{\bm r}_{jj'})}{r_{jj'}^5}\right],
\label{seq:dipolar0}
\ee
where $\mu_{\rm B}$ is the Bohr magneton, and ${\bm r}_{jj'}$ is the relative distance between spins $j$ and $j'$. It is important to consider the combined effect of the Zeeman term, 
\be
\hat{ H}_{\rm z} = \Delta\sum_i {\hat S}_i^z,
\ee
and dipolar interactions. In particular, in the presence of a Zeeman field, it is convenient to pick a quantization axis which is canted from the 2D plane ${\bm r} = (x,y,0)$, 
\be
{\hat S}_{j}^z \rightarrow \cos\theta {\hat S}_{j}^z - \sin\theta {\hat S}_j^x, \quad {\hat S}_j^x \rightarrow \cos\theta {\hat S}_j^x + \sin\theta {\hat S}_j^z, \quad {\hat S}_j^y \rightarrow {\hat S}_j^y, 
\label{seq:rotation}
\ee
where $\theta$ will be conveniently chosen below. Inserting Eq.\,(\ref{seq:rotation}) into Eq.\,(\ref{seq:dipolar0}), we find
\be
\begin{array}{rr}
\displaystyle \hat{ H}_{\rm d} = & \displaystyle \frac{\mu_{\rm B}^2}{8\pi} \sum_{j\tau} \frac{1}{{\tau}^3} \left[ {\hat S}_{j}^x{\hat S}_{j+\tau}^x\left(1-3\cos^2\theta\frac{\tau_{x}^2}{{\tau}^2}\right) + {\hat S}_{j}^y{\hat S}_{j+\tau}^y\left(1-3\frac{\tau_{y}^2}{{\tau}^2}\right) + {\hat S}_{j}^z{\hat S}_{j+\tau}^z\left(1-3\sin^2\theta\frac{\tau_{x}^2}{{\tau}^2}\right) \right.\\ & \\
& \displaystyle \left. - 6 \sin\theta\cos\theta\frac{\tau_x^2}{{\tau}^2}{\hat S}_{j}^x{\hat S}_{j+\tau}^z - 6 \cos\theta\frac{\tau_x\tau_y}{{\tau}^2}{\hat S}_{j}^x{\hat S}_{j+\tau}^y - 6 \sin\theta\frac{\tau_x\tau_y}{{\tau}^2}{\hat S}_{j}^z{\hat S}_{j+\tau}^y\right],
\end{array}
\ee
where ${\tau}$ denotes relative positions between spins on a two-dimensional square lattice (not restricted to nearest neighbors). After rearranging terms, we find 
\be
\begin{array}{rr}
\displaystyle \hat{ H}_{\rm d} = & \displaystyle \frac{3\mu_{\rm B}^2}{8\pi} \sum_{j\tau} \frac{1}{{\tau}^3} \left[ \left(\hat{\bm S}_{j}\cdot\hat{\bm S}_{j+\tau}\right)\left(\frac{1}{3}-\frac{\tau_{x}^2}{{\tau}^2}\right) + \sin^2\theta \frac{\tau_{x}^2}{{\tau}^2} {\hat S}_{j}^x{\hat S}_{j+\tau}^x +  \cos^2\theta \frac{\tau_{x}^2}{{\tau}^2} {\hat S}_{j}^z{\hat S}_{j+\tau}^z - \frac{\tau_y^2 - \tau_x^2}{{\tau}^5}{\hat S}_j^y{\hat S}_{j+\tau}^y\right.\\ & \\
& \displaystyle \left. - 2 \sin\theta\cos\theta\frac{\tau_x^2}{{\tau}^2}{\hat S}_{j}^x{\hat S}_{j+\tau}^z - 2 \cos\theta\frac{\tau_x\tau_y}{{\tau}^2}{\hat S}_{j}^x{\hat S}_{j+\tau}^y - 2 \sin\theta\frac{\tau_x\tau_y}{{\tau}^2}{\hat S}_{j}^z{\hat S}_{j+\tau}^y\right].
\end{array}
\ee
Note that the first term on the right-hand side can be incorporated into the definition of $J$ with a small anisotropy in the $x$ direction which we will neglect. For convenience, we define $\hat{ H}_{\rm d} = \hat{ H}_{zz} + \hat{ H}_{xz}+\hat{ H}_{xx} + \hat{ H}_{yy} +\hat{ H}_{xy} + \hat{ H}_{yz}$, with 
\be
\begin{array}{c}
\displaystyle\hat{ H}_{zz} = \varepsilon_{\rm d}\cos^2\theta \frac{a^3}{\pi S^2}\sum_{j\tau} \frac{\tau_x^2}{{\tau}^5}{\hat S}_j^z{\hat S}_{j+\tau}^z ,\,\, \hat{ H}_{xx} = \varepsilon_{\rm d}\sin^2\theta \frac{a^3}{\pi S^2}\sum_{j\tau} \frac{\tau_x^2}{\tau^5}{\hat S}_j^x{\hat S}_{j+\tau}^x, \,\, \hat{ H}_{xz} = - 2 \varepsilon_{\rm d} \sin\theta\cos\theta \frac{a^3}{\pi S^2}\sum_{j\tau} \frac{\tau_x\tau_y}{\tau^5}{\hat S}_j^x{\hat S}_{j+\tau}^z, \\
\displaystyle \hat{ H}_{yy} = \varepsilon_{\rm d}\frac{a^3}{\pi S^2}\sum_{j\tau} \frac{\tau_y^2 - \tau_x^2}{{\tau}^5}{\hat S}_j^y{\hat S}_{j+\tau}^y,\, \hat{ H}_{xy} = -2\varepsilon_{\rm d} \cos\theta \frac{a^3}{\pi S^2}\sum_{j\tau} \frac{\tau_x\tau_y}{\tau^5}{\hat S}_j^x{\hat S}_{j+\tau}^y, \quad \hat{ H}_{yz} = -2\varepsilon_{\rm d} \sin\theta \frac{a^3}{\pi S^2}\sum_{j\tau} \frac{\tau_x\tau_y}{\tau^5}{\hat S}_j^y{\hat S}_{j+\tau}^z,
\end{array}
\label{seq:xxzzodd}
\ee
where we defined the dipolar energy as
\be
\varepsilon_{\rm d} = \frac{3 S^2\mu_{\rm B}^2}{4a^3}.
\ee

The Zeeman splitting term in the rotated frame is given $\hat{ H}_{\rm z} = \hat{ H}_{x}+\hat{ H}_z$, with 
\be
\hat{ H}_{x} = \Delta \cos\theta \sum_j  {\hat S}_j^x, \quad \hat{ H}_{z} =  -\Delta \sin\theta \sum_j  {\hat S}_j^z.
\ee

Focusing on $\hat{ H}_{zz}$ first, we define ${\hat S}_j^z = -S(1-{\hat n}_j)$, which leads to
\be
\hat{ H}_{zz} = \varepsilon_{\rm d} \cos^2\theta \frac{a^3}{\pi}\sum_{j\tau}\frac{\tau_x^2}{{\tau}^5}(1-2{\hat n}_j+{\hat n}_j{\hat n}_{j+\tau})=\varepsilon_{\rm d}\cos^2\theta \left( NS - 2\sum_j{\hat n}_j + \frac{a^3}{\pi}\sum_{j\tau}\frac{\tau_x^2}{{\tau}^5}{\hat n}_j{\hat n}_{j+\tau}\right), 
\ee
and where, in the last step, we used 
\be
\sum_{\boldsymbol\tau}e^{-i{\bm k}\cdot{\boldsymbol\tau}}\frac{\tau_x^2}{{\tau}^5} = \frac{\pi}{a^3} + { O}(q^2).
\label{seq:sumtau}
\ee

Similarly, for $\hat{ H}_{xz}$  we find
\be
\hat{ H}_{xz} = 2\varepsilon_{\rm d}\sin\theta\cos\theta \frac{a^3}{\pi S}\sum_{j\tau}\frac{\tau_x^2}{{\tau}^5}{\hat S}_j^x(1-{\hat n}_{j+\tau})=\frac{2\varepsilon_{\rm d}\sin\theta\cos\theta}{S}\left(\sum_j{\hat S}_j^x - \frac{a^3}{\pi}\sum_{j\tau}\frac{\tau_x^2}{{\tau}^5}{\hat S}_j^x{\hat n}_{j+\tau}\right).
\ee
Turning to $\hat{ H}_{xx}$ and using ${\hat S}_j^x = ({\hat S}_j^++{\hat S}_j^-)/2$, we find
\be
\hat{ H}_{xx} = \frac{\varepsilon_{\rm d}\sin^2\theta}{4}\frac{a^3}{\pi S^2}\sum_{j\tau}\frac{\tau_x^2}{{\tau}^5}\left({\hat S}_j^+{\hat S}_{j+\tau}^++{\hat S}_j^-{\hat S}_{j+\tau}^-+2{\hat S}_j^+{\hat S}_{j+\tau}^-\right) = \frac{\varepsilon_{\rm d}\sin^2\theta}{4S^2}\sum_{\bm k}\left({\hat S}_{-\bm k}^+{\hat S}_{\bm k}^++{\hat S}_{-\bm k}^-{\hat S}_{\bm k}^-+2{\hat S}_{-\bm k}^+{\hat S}_{\bm k}^-\right),
\ee
where, in the last step, we used Eq.(\ref{seq:sumtau}). The term $\hat{ H}_{xx}$ introduces coherent creation/destruction of two magnons. The term $\hat{ H}_{xy}$ also introduces similar two-magnon processes such as those in $\hat{ H}_{xx}$, 
\be
\hat{ H}_{xy} = -\frac{\varepsilon_{\rm d}\cos\theta}{2i}\frac{a^3}{\pi S^2}\sum_{j\tau}\frac{\tau_x\tau_y}{|{\boldsymbol\tau}|^5}\left({\hat S}_{j}^+{\hat S}_{j+\tau}^+ - {\hat S}_{j}^-{\hat S}_{j+\tau}^-\right) = -\frac{2\varepsilon_{\rm d}\cos\theta}{i\pi S^2}\sum_{\bm k}\frac{k_xk_y}{a}\left({\hat S}_{-\bm k}^+{\hat S}_{\bm k}^+-{\hat S}_{-\bm k}^-{\hat S}_{\bm k}^-\right),
\label{seq:Hxy1}
\ee
but the matrix elements of $\hat{ H}_{xy}$ are ${ O}(q^2)$ smaller than those corresponding to $\hat{ H}_{xx}$ [in the last step of Eq.(\ref{seq:Hxy1}), we used $\sum_{\boldsymbol\tau} e^{i{\bm k}\cdot {\boldsymbol\tau}} \frac{\tau_x\tau_y}{{\tau}^5} =  \frac{4 k_xk_y}{a} + { O}(k^4)$]. As a result, we neglect $\hat{ H}_{xy}$. Finally, for $\hat{ H}_{yz}$, we find
\be
\begin{array}{rl}
\displaystyle\hat{ H}_{yz} = & \displaystyle -2\varepsilon_{\rm d}\sin\theta\frac{a^3}{\pi S}\sum_{j\tau}\frac{\tau_x\tau_y}{{\tau}^5}{\hat S}_j^y (1-n_{j+\tau}) = -2\varepsilon_{\rm d}\sin\theta\frac{a^3}{\pi S}\left(\sum_{j\tau}\frac{\tau_x\tau_y}{{\tau}^5}{\hat S}_j^y - \sum_{j\tau}\frac{\tau_x\tau_y}{{\tau}^5}{\hat S}_j^y{\hat n}_{j+\tau}\right) \\
\displaystyle = & \displaystyle 6\varepsilon_{\rm d}\sin\theta\frac{a^3}{\pi S}\sum_{j\tau}\frac{\tau_x\tau_y}{{\tau}^5}{\hat S}_j^y{\hat n}_{j+\tau},
\end{array}
\ee
where the first term in the third equality is zero because $\sum_{\boldsymbol\tau} \tau_x\tau_y / {\tau}^5 = 0$, thus giving only a cubic term. The cubic term, however, has matrix elements ${ O}(q^2)$ smaller than those corresponing to $\hat{ H}_{xz}$ because of the factors $\tau_x\tau_y$. As a result, we neglect the matrix elements introduced by $\hat{ H}_{yz}$ when compared to those in $\hat{ H}_{xz}$.

The Zeeman splitting term $\hat{ H}_x$ and the dipolar term $\hat{ H}_{xz}$ both generate terms which are linear in ${\hat S}_i^x$. In particular, 
\be
\hat{ H}_{x}+\hat{ H}_{xz} = -\Delta\sin\theta\sum_{j}{\hat S}_j^x + \frac{2\varepsilon_{\rm d}\sin\theta\cos\theta}{S}\sum_j{\hat S}_j^x - \frac{2\varepsilon_{\rm d}\sin\theta\cos\theta}{S}\sum_{j\tau}\frac{\tau_x^2}{{\tau}^5}{\hat S}_j^x{\hat n}_{j+\tau}. 
\ee
As a result, we conveniently define $\theta$ such that the linear term is cancelled. This leads to 
\be
\begin{array}{cc}
\displaystyle\cos\theta = \frac{S\Delta}{2\varepsilon_{\rm d}},\quad\quad  & 0\le S\Delta\le 2 \varepsilon_{\rm d}, \\ 
\displaystyle\theta = 0,\quad\quad & S\Delta> 2 \varepsilon_{\rm d}.
\end{array}
\label{seq:thetamin}
\ee
Therefore, in this case, the terms 
\be
\hat{ H}_{x}+\hat{ H}_{xz} = -\frac{2\varepsilon_{\rm d}\sin\theta\cos\theta}{S}\sum_{j\tau}\frac{\tau_x^2}{{\tau}^5}{\hat S}_j^x{\hat n}_{j+\tau}, 
\label{seq:Hx+Hxz}
\ee
lead to a cubic interaction term after a Holstein-Primakoff transformation.  

In the same spirit, combining $\hat{ H}_z$ from Zeeman splitting and $\hat{ H}_{zz}$ from dipolar interaction, we find
\be
\hat{ H}_{z}+\hat{ H}_{zz} = \left(\Delta S\cos\theta-2\varepsilon_{\rm d}\cos^2\theta\right)\sum_j {\hat n}_j + \varepsilon_{\rm d}\cos^2\theta\frac{a^3}{\pi} \sum_{j\tau}\frac{\tau_x^2}{\tau^5}{\hat n}_j{\hat n}_{j+\tau}. 
\label{seq:Hz+Hzz}
\ee
As a result, the combination of ${ H}_{z}$ and ${ H}_{zz}$ gives rise to a magnon gap induced by Zeeman splitting and dipolar interactions, and a quartic interaction induced by dipolar interactions. 

\subsection{Effective Hamiltonian}

To cast the dipolar Hamiltonian into a long-wavelength, effective Hamiltonian, we use the Holstein-Primakoff transformation to leading order, which results in 
\be
\sum_{j\tau}\frac{\tau_x^2}{{\tau}^5}{\hat n}_j{\hat n}_{j+\tau} = \sum_{j\tau}\frac{\tau_x^2}{{\tau}^5}{\hat a}_j^\dagger {\hat a}_{j+\tau}^\dagger {\hat a}_{j+\tau}{\hat a}_j = \sum_{\bm k \bm p \bm q}\left(\sum_{\tau}e^{-i{\bm q}\cdot{\boldsymbol\tau}}\frac{\tau_x^2}{{\tau}^5}\right){\hat a}_{\bm k + \bm q}^\dagger {\hat a}_{\bm p - \bm q}^\dagger {\hat a}_{\bm p}{\hat a}_{\bm k} \approx \frac{\pi}{a^3}\sum_{\bm k \bm p \bm q}{\hat a}_{\bm k + \bm q}^\dagger {\hat a}_{\bm p - \bm q}^\dagger {\hat a}_{\bm p}{\hat a}_{\bm k}.
\ee
In the last step, we used Eq.(\ref{seq:sumtau}). In addition, for Eq.(\ref{seq:Hx+Hxz}), we use
\be
\begin{array}{r}
\displaystyle \sum_{j\tau}\frac{\tau_x^2}{{\tau}^5}{\hat S}_j^x{\hat n}_{j+\tau} = \sqrt{\frac{S}{2}}\sum_{j\tau}\frac{\tau_x^2}{{\tau}^5}\left({\hat a}_j^\dagger {\hat a}_{j+\tau}^\dagger {\hat a}_{j+\tau} + {\hat a}_{j+\tau}^\dagger {\hat a}_{j+\tau}{\hat a}_j\right) = \sqrt{\frac{S}{2N}} \sum_{\bm k \bm p {\boldsymbol\tau}}\frac{\tau_x^2}{{\tau}^5}\left[e^{-{\bm p}\cdot{\boldsymbol\tau}}{\hat a}_{\bm p}^\dagger {\hat a}_{\bm k}^\dagger {\hat a}_{\bm k + \bm p} + e^{-i{\bm k}\cdot p}{\hat a}_{\bm k+\bm p}^\dagger {\hat a}_{\bm p}{\hat a}_{\bm k}\right]\\
\displaystyle \approx \sqrt{\frac{S}{2N}}\frac{\pi}{a^3}\sum_{\bm k \bm p}\left({\hat a}_{\bm p}^\dagger {\hat a}_{\bm k}^\dagger {\hat a}_{\bm k + \bm p} + {\hat a}_{\bm k + \bm p}^\dagger {\hat a}_{\bm p}{\hat a}_{\bm k}\right).
\end{array}
\ee
Putting everything together, we find that, at long wavelength, the dipolar and Zeeman Hamiltonian can be effectively written as 
\be
\begin{array}{c}
\displaystyle\hat{ H}_{\rm d} + \hat{ H}_{\rm z} \approx \sum_{\bm k} \left[ \Delta{\hat a}_{\bm k}^\dagger {\hat a}_{\bm k} + \lambda_2\left( {\hat a}_{\bm k}{\hat a}_{-\bm k} + {\hat a}_{\bm k}^\dagger {\hat a}_{-\bm k}^\dagger\right) \right] - \frac{\lambda_3}{\sqrt{N}}\sum_{\bm k \bm p}\left({\hat a}_{\bm p}^\dagger {\hat a}_{\bm k}^\dagger {\hat a}_{\bm k + \bm p} +  {\hat a}_{\bm k + \bm p}^\dagger {\hat a}_{\bm p} {\hat a}_{\bm k} \right) +  \frac{\lambda_4}{N}\sum_{{\bm k}{\bm p}{\bm q}}a_{{\bm p}+{\bm q}}^\dagger {\hat a}_{{\bm k}-{\bm q}}^\dagger {\hat a}_{{\bm p}}{\hat a}_{{\bm k}},\\ \\
\displaystyle \tilde\Delta = \left(\Delta S\cos\theta-2\varepsilon_{\rm d}\cos^2\theta\right)+\frac{\varepsilon_{\rm d}\sin^2\theta}{S},\quad \lambda_2= \frac{\varepsilon_{\rm d}\sin^2\theta}{2S}, \quad \lambda_3 = \varepsilon_{\rm d}\sqrt{2/S}\sin\theta\cos\theta, \quad \lambda_4 = \varepsilon_{\rm d}\cos^2\theta.
\end{array}
\label{seq:Heven}
\ee

\subsection{Bogoliubov transformation}

For small Zeeman fields, the canting angle lies in the range $0< \theta \le \pi/2$, and $\lambda_{2,3}$ are finite. The quadratic part of the Heisenberg Hamiltonian combined with Eq.(\ref{seq:Heven}),
\be
\hat{ H}_2 = \sum_{\bm k}\left[(\Delta + \varepsilon_{\bm k}){\hat a}_{\bm k}^\dagger {\hat a}_{\bm k} + \lambda_2 ({\hat a}_{\bm k}{\hat a}_{-\bm k} + {\hat a}_{\bm k}^\dagger {\hat a}_{-\bm k}^\dagger)\right], 
\ee
can be diagonalized using a Bogoliubov transformation:
\be
{\hat a}_{\bm k} = s_{\bm k} {\hat\beta}_{\bm k} + t_{\bm k}{\hat\gamma}_{-\bm k}^\dagger,\quad {\hat a}_{-\bm k} = s_{\bm k} {\hat\beta}_{\bm k} + t_{\bm k}{\hat\gamma}_{-\bm k}^\dagger,
\ee
where $s_{\bm k}$ and $t_{\bm k}$ are ${\bm k}$-dependent real numbers. It is straightforward to show that 
\be
s_{\bm k} = \cosh\varphi_{\bm k}, \quad t_{\bm k} = \sinh\varphi_{\bm k}, 
\ee
diagonalizes $\hat{ H}_2$, 
\be
\hat{ H}_2 = \sum_{\bm k} E_{\bm k} \left[\beta_{\bm k}^\dagger\beta_{\bm k} + \gamma_{\bm k}^\dagger\gamma_{\bm k}\right], \quad E_{\bm k} = \sqrt{(\varepsilon_{\bm k} + \Delta)^2 - \lambda_2^2},
\ee
where $\varphi_{\bm k}$ is the solution of 
\be
\sinh 2\varphi_{\bm k} = -\frac{\lambda_2}{2E_{\bm k}}.
\ee

Several comments are in order. First, we note that the magnon dispersion is quadratic, with or without dipolar interactions. In particular, in the presence of dipolar interactions, there will be a small correction to the magnon mass at low energies on the order of ${\cal O}(\varepsilon_{\rm d} / J)$, and which we will neglect (quadratic dispersion greatly simplifies the hydrodynamic description, as will be discussed below). Second, we are mainly interested on the hydrodynamic behavior at large $T$ such that magnon-magnon collisions become important. In the regime $\varepsilon_{\rm d} \ll T \ll J$, most magnons will typically have large kinetic energies $\varepsilon_{\bm k}$ such that corrections due to the Bogoliubov transformation are negligible. 

For sufficiently large Zeeman fields, when $\Delta\ge \varepsilon_{\rm d}$ and $\theta = 0 $, then the coupling terms verigy $\lambda_{2,3} = 0$. In this case, the quadratic part of $\hat{ H}_J+\hat{ H}_{\rm d} + \hat{ H}_{\rm z}$ is already diagonal in the $({\hat a}_{\bm k},a_{\bm k}^\dagger)$ basis and there is no need for a Bogoliubov transformation. 

\subsection{Magnon leakage}

Three magnon processes in Eq.(\ref{seq:Heven}) do not preserve particle number. This means that the distribution function $\bar{n}_{\bm k} = [z^{-1}e^{\varepsilon_{\bm k}/T}-1]^{-1}$ is a quasi-equilibrium distribution if $0<z<1$, and invalidates our hydrodynamic theory for frequencies below the leakage rate. The total magnon leakage rate can be calculated from 
\be
\frac{dn}{dt} = - \frac{\lambda_3^2}{N^2}\sum_{\bm k \bm p}2\pi\delta(\varepsilon_{\bm k} + \varepsilon_{\bm p}+\Delta - \varepsilon_{\bm k + \bm p})\left[\bar{n}_{\bm k}\bar{n}_{\bm p}(1+\bar{n}_{\bm k + \bm p})-(1+\bar{n}_{\bm k})(1+\bar{n}_{\bm p})\bar{n}_{\bm k + \bm p}\right].
\ee
Here we note that three magnon processes are not necessarily suppressed by energy and momenta conservation. For instance, if the incoming magnon states have momenta that verifies ${\bm k}\cdot{\bm p} = m \Delta$, then energy and momentum is conserved after the collision. For concreteness, let us assume that $ u_\alpha \ll \sqrt{T/m}$, which leads to 
\be
\frac{dn}{dt} = - \frac{{\gamma}_{\rm leak}(z^2-z^3)}{4\pi}\frac{T \lambda_3^2}{J^2a^2}, \quad {\gamma}_{\rm leak} = \frac{16\pi}{z^3}\int\frac{d\tilde{\bm k}}{(2\pi)^2}\int\frac{d\tilde{\bm p}}{(2\pi)^2}\delta[\tilde{\bm k}^2 + \tilde{\bm p}^2 +\tilde\Delta- (\tilde{\bm k} + \tilde{\bm p})^2]\bar{n}_{\tilde{\bm k}}\bar{n}_{\tilde{\bm p}}\bar{n}_{\tilde{\bm k} + \tilde{\bm p}} e^{\tilde{\bm k}^2 + \tilde{\bm p}^2},
\ee
where we normalized $\tilde{\bm k} = \bm k / \bar{k}$. The value of ${\gamma}_{\rm leak}(z)$ can be shown numerically to be  $\gamma_{\rm leak} \sim {\cal O}(1)$. From here we can define the leakage rate
\be
\frac{1}{\tau_{\rm leak}} = \frac{1}{n}\frac{dn}{dt}= \frac{{\gamma}_{\rm leak}\lambda_3^2}{2J}(z^2-z^3).
\ee
Using $J\sim 1000\,{\rm K}$, $\lambda_3 \sim 1\,{\rm K}$, and $z\approx 0.9$, we obtain $1/\tau_{\rm leak} \sim 5\,{\rm MHz}$. As such, magnon number can be assumed to be a good conserved quantity for $\omega\gg 1\,{\rm MHz}$. 

\end{widetext}


\begin{thebibliography}{40}%
\makeatletter
\providecommand \@ifxundefined [1]{%
 \@ifx{#1\undefined}
}%
\providecommand \@ifnum [1]{%
 \ifnum #1\expandafter \@firstoftwo
 \else \expandafter \@secondoftwo
 \fi
}%
\providecommand \@ifx [1]{%
 \ifx #1\expandafter \@firstoftwo
 \else \expandafter \@secondoftwo
 \fi
}%
\providecommand \natexlab [1]{#1}%
\providecommand \enquote  [1]{``#1''}%
\providecommand \bibnamefont  [1]{#1}%
\providecommand \bibfnamefont [1]{#1}%
\providecommand \citenamefont [1]{#1}%
\providecommand \href@noop [0]{\@secondoftwo}%
\providecommand \href [0]{\begingroup \@sanitize@url \@href}%
\providecommand \@href[1]{\@@startlink{#1}\@@href}%
\providecommand \@@href[1]{\endgroup#1\@@endlink}%
\providecommand \@sanitize@url [0]{\catcode `\\12\catcode `\$12\catcode
  `\&12\catcode `\#12\catcode `\^12\catcode `\_12\catcode `\%12\relax}%
\providecommand \@@startlink[1]{}%
\providecommand \@@endlink[0]{}%
\providecommand \url  [0]{\begingroup\@sanitize@url \@url }%
\providecommand \@url [1]{\endgroup\@href {#1}{\urlprefix }}%
\providecommand \urlprefix  [0]{URL }%
\providecommand \Eprint [0]{\href }%
\providecommand \doibase [0]{http://dx.doi.org/}%
\providecommand \selectlanguage [0]{\@gobble}%
\providecommand \bibinfo  [0]{\@secondoftwo}%
\providecommand \bibfield  [0]{\@secondoftwo}%
\providecommand \translation [1]{[#1]}%
\providecommand \BibitemOpen [0]{}%
\providecommand \bibitemStop [0]{}%
\providecommand \bibitemNoStop [0]{.\EOS\space}%
\providecommand \EOS [0]{\spacefactor3000\relax}%
\providecommand \BibitemShut  [1]{\csname bibitem#1\endcsname}%
\let\auto@bib@innerbib\@empty
\bibitem [{\citenamefont {Torre}\ \emph {et~al.}(2015)\citenamefont {Torre},
  \citenamefont {Tomadin}, \citenamefont {Geim},\ and\ \citenamefont
  {Polini}}]{2015polini}%
  \BibitemOpen
  \bibfield  {author} {\bibinfo {author} {\bibfnamefont {I.}~\bibnamefont
  {Torre}}, \bibinfo {author} {\bibfnamefont {A.}~\bibnamefont {Tomadin}},
  \bibinfo {author} {\bibfnamefont {A.~K.}\ \bibnamefont {Geim}}, \ and\
  \bibinfo {author} {\bibfnamefont {M.}~\bibnamefont {Polini}},\ }\href@noop {}
  {\bibfield  {journal} {\bibinfo  {journal} {Phys. Rev. B}\ }\textbf {\bibinfo
  {volume} {92}},\ \bibinfo {pages} {165433} (\bibinfo {year}
  {2015})}\BibitemShut {NoStop}%
\bibitem [{\citenamefont {Bandurin}\ \emph {et~al.}(2016)\citenamefont
  {Bandurin}, \citenamefont {Torre}, \citenamefont {Kumar}, \citenamefont
  {Ben~Shalom}, \citenamefont {Tomadin}, \citenamefont {Principi},
  \citenamefont {Auton}, \citenamefont {Khestanova}, \citenamefont {Novoselov},
  \citenamefont {Grigorieva}, \citenamefont {Ponomarenko}, \citenamefont
  {Geim},\ and\ \citenamefont {Polini}}]{2016bandurin}%
  \BibitemOpen
  \bibfield  {author} {\bibinfo {author} {\bibfnamefont {D.~A.}\ \bibnamefont
  {Bandurin}}, \bibinfo {author} {\bibfnamefont {I.}~\bibnamefont {Torre}},
  \bibinfo {author} {\bibfnamefont {R.~K.}\ \bibnamefont {Kumar}}, \bibinfo
  {author} {\bibfnamefont {M.}~\bibnamefont {Ben~Shalom}}, \bibinfo {author}
  {\bibfnamefont {A.}~\bibnamefont {Tomadin}}, \bibinfo {author} {\bibfnamefont
  {A.}~\bibnamefont {Principi}}, \bibinfo {author} {\bibfnamefont {G.~H.}\
  \bibnamefont {Auton}}, \bibinfo {author} {\bibfnamefont {E.}~\bibnamefont
  {Khestanova}}, \bibinfo {author} {\bibfnamefont {K.~S.}\ \bibnamefont
  {Novoselov}}, \bibinfo {author} {\bibfnamefont {I.~V.}\ \bibnamefont
  {Grigorieva}}, \bibinfo {author} {\bibfnamefont {L.~A.}\ \bibnamefont
  {Ponomarenko}}, \bibinfo {author} {\bibfnamefont {A.~K.}\ \bibnamefont
  {Geim}}, \ and\ \bibinfo {author} {\bibfnamefont {M.}~\bibnamefont
  {Polini}},\ }\href@noop {} {\bibfield  {journal} {\bibinfo  {journal}
  {Science}\ }\textbf {\bibinfo {volume} {351}},\ \bibinfo {pages} {1055}
  (\bibinfo {year} {2016})}\BibitemShut {NoStop}%
\bibitem [{\citenamefont {Crossno}\ \emph {et~al.}(2016)\citenamefont
  {Crossno}, \citenamefont {Shi}, \citenamefont {Wang}, \citenamefont {Liu},
  \citenamefont {Harzheim}, \citenamefont {Lucas}, \citenamefont {Sachdev},
  \citenamefont {Kim}, \citenamefont {Taniguchi}, \citenamefont {Watanabe},
  \citenamefont {Ohki},\ and\ \citenamefont {Fong}}]{2016crossno}%
  \BibitemOpen
  \bibfield  {author} {\bibinfo {author} {\bibfnamefont {J.}~\bibnamefont
  {Crossno}}, \bibinfo {author} {\bibfnamefont {J.~K.}\ \bibnamefont {Shi}},
  \bibinfo {author} {\bibfnamefont {K.}~\bibnamefont {Wang}}, \bibinfo {author}
  {\bibfnamefont {X.}~\bibnamefont {Liu}}, \bibinfo {author} {\bibfnamefont
  {A.}~\bibnamefont {Harzheim}}, \bibinfo {author} {\bibfnamefont
  {A.}~\bibnamefont {Lucas}}, \bibinfo {author} {\bibfnamefont
  {S.}~\bibnamefont {Sachdev}}, \bibinfo {author} {\bibfnamefont
  {P.}~\bibnamefont {Kim}}, \bibinfo {author} {\bibfnamefont {T.}~\bibnamefont
  {Taniguchi}}, \bibinfo {author} {\bibfnamefont {K.}~\bibnamefont {Watanabe}},
  \bibinfo {author} {\bibfnamefont {T.~A.}\ \bibnamefont {Ohki}}, \ and\
  \bibinfo {author} {\bibfnamefont {K.~C.}\ \bibnamefont {Fong}},\ }\href@noop
  {} {\bibfield  {journal} {\bibinfo  {journal} {Science}\ }\textbf {\bibinfo
  {volume} {351}},\ \bibinfo {pages} {1058} (\bibinfo {year}
  {2016})}\BibitemShut {NoStop}%
\bibitem [{\citenamefont {Levitov}\ and\ \citenamefont
  {Falkovich}(2016)}]{2016levitovfalkovich}%
  \BibitemOpen
  \bibfield  {author} {\bibinfo {author} {\bibfnamefont {L.}~\bibnamefont
  {Levitov}}\ and\ \bibinfo {author} {\bibfnamefont {G.}~\bibnamefont
  {Falkovich}},\ }\href@noop {} {\bibfield  {journal} {\bibinfo  {journal}
  {Nature Physics}\ }\textbf {\bibinfo {volume} {12}},\ \bibinfo {pages} {672}
  (\bibinfo {year} {2016})}\BibitemShut {NoStop}%
\bibitem [{\citenamefont {Guo}\ \emph {et~al.}(2017)\citenamefont {Guo},
  \citenamefont {Ilseven}, \citenamefont {Falkovich},\ and\ \citenamefont
  {Levitov}}]{2017superballisticconduction}%
  \BibitemOpen
  \bibfield  {author} {\bibinfo {author} {\bibfnamefont {H.}~\bibnamefont
  {Guo}}, \bibinfo {author} {\bibfnamefont {E.}~\bibnamefont {Ilseven}},
  \bibinfo {author} {\bibfnamefont {G.}~\bibnamefont {Falkovich}}, \ and\
  \bibinfo {author} {\bibfnamefont {L.~S.}\ \bibnamefont {Levitov}},\
  }\href@noop {} {\bibfield  {journal} {\bibinfo  {journal} {Proceedings of the
  National Academy of Sciences}\ }\textbf {\bibinfo {volume} {114}},\ \bibinfo
  {pages} {3068} (\bibinfo {year} {2017})}\BibitemShut {NoStop}%
\bibitem [{\citenamefont {Krishna~Kumar}\ \emph {et~al.}(2017)\citenamefont
  {Krishna~Kumar}, \citenamefont {Bandurin}, \citenamefont {Pellegrino},
  \citenamefont {Cao}, \citenamefont {Principi}, \citenamefont {Guo},
  \citenamefont {Auton}, \citenamefont {Ben~Shalom}, \citenamefont
  {Ponomarenko}, \citenamefont {Falkovich}, \citenamefont {Watanabe},
  \citenamefont {Taniguchi}, \citenamefont {Grigorieva}, \citenamefont
  {Levitov}, \citenamefont {Polini},\ and\ \citenamefont
  {Geim}}]{2017superballisticflowexp}%
  \BibitemOpen
  \bibfield  {author} {\bibinfo {author} {\bibfnamefont {R.}~\bibnamefont
  {Krishna~Kumar}}, \bibinfo {author} {\bibfnamefont {D.~A.}\ \bibnamefont
  {Bandurin}}, \bibinfo {author} {\bibfnamefont {F.~M.~D.}\ \bibnamefont
  {Pellegrino}}, \bibinfo {author} {\bibfnamefont {Y.}~\bibnamefont {Cao}},
  \bibinfo {author} {\bibfnamefont {A.}~\bibnamefont {Principi}}, \bibinfo
  {author} {\bibfnamefont {H.}~\bibnamefont {Guo}}, \bibinfo {author}
  {\bibfnamefont {G.}~\bibnamefont {Auton}}, \bibinfo {author} {\bibfnamefont
  {M.}~\bibnamefont {Ben~Shalom}}, \bibinfo {author} {\bibfnamefont {L.~A.}\
  \bibnamefont {Ponomarenko}}, \bibinfo {author} {\bibfnamefont
  {G.}~\bibnamefont {Falkovich}}, \bibinfo {author} {\bibfnamefont
  {K.}~\bibnamefont {Watanabe}}, \bibinfo {author} {\bibfnamefont
  {T.}~\bibnamefont {Taniguchi}}, \bibinfo {author} {\bibfnamefont
  {I.}~\bibnamefont {Grigorieva}}, \bibinfo {author} {\bibfnamefont {L.~S.}\
  \bibnamefont {Levitov}}, \bibinfo {author} {\bibfnamefont {M.}~\bibnamefont
  {Polini}}, \ and\ \bibinfo {author} {\bibfnamefont {A.}~\bibnamefont
  {Geim}},\ }\href@noop {} {\bibfield  {journal} {\bibinfo  {journal} {Nature
  Physics}\ }\textbf {\bibinfo {volume} {13}},\ \bibinfo {pages} {1182}
  (\bibinfo {year} {2017})}\BibitemShut {NoStop}%
\bibitem [{\citenamefont {Davison}\ \emph {et~al.}(2016)\citenamefont
  {Davison}, \citenamefont {Delacr\'etaz}, \citenamefont {Gout\'eraux},\ and\
  \citenamefont {Hartnoll}}]{2016superconductivity}%
  \BibitemOpen
  \bibfield  {author} {\bibinfo {author} {\bibfnamefont {R.~A.}\ \bibnamefont
  {Davison}}, \bibinfo {author} {\bibfnamefont {L.~V.}\ \bibnamefont
  {Delacr\'etaz}}, \bibinfo {author} {\bibfnamefont {B.}~\bibnamefont
  {Gout\'eraux}}, \ and\ \bibinfo {author} {\bibfnamefont {S.~A.}\ \bibnamefont
  {Hartnoll}},\ }\href@noop {} {\bibfield  {journal} {\bibinfo  {journal}
  {Phys. Rev. B}\ }\textbf {\bibinfo {volume} {94}},\ \bibinfo {pages} {054502}
  (\bibinfo {year} {2016})}\BibitemShut {NoStop}%
\bibitem [{\citenamefont {Lucas}(2016)}]{2016lucassoundmodes}%
  \BibitemOpen
  \bibfield  {author} {\bibinfo {author} {\bibfnamefont {A.}~\bibnamefont
  {Lucas}},\ }\href@noop {} {\bibfield  {journal} {\bibinfo  {journal} {Phys.
  Rev. B}\ }\textbf {\bibinfo {volume} {93}},\ \bibinfo {pages} {245153}
  (\bibinfo {year} {2016})}\BibitemShut {NoStop}%
\bibitem [{\citenamefont {Lucas}\ \emph {et~al.}(2016)\citenamefont {Lucas},
  \citenamefont {Davison},\ and\ \citenamefont {Sachdev}}]{2016lucas-weyl}%
  \BibitemOpen
  \bibfield  {author} {\bibinfo {author} {\bibfnamefont {A.}~\bibnamefont
  {Lucas}}, \bibinfo {author} {\bibfnamefont {R.~A.}\ \bibnamefont {Davison}},
  \ and\ \bibinfo {author} {\bibfnamefont {S.}~\bibnamefont {Sachdev}},\
  }\href@noop {} {\bibfield  {journal} {\bibinfo  {journal} {Proceedings of the
  National Academy of Sciences}\ }\textbf {\bibinfo {volume} {113}},\ \bibinfo
  {pages} {9463} (\bibinfo {year} {2016})}\BibitemShut {NoStop}%
\bibitem [{\citenamefont {Castro-Alvaredo}\ \emph {et~al.}(2016)\citenamefont
  {Castro-Alvaredo}, \citenamefont {Doyon},\ and\ \citenamefont
  {Yoshimura}}]{integrable1}%
  \BibitemOpen
  \bibfield  {author} {\bibinfo {author} {\bibfnamefont {O.~A.}\ \bibnamefont
  {Castro-Alvaredo}}, \bibinfo {author} {\bibfnamefont {B.}~\bibnamefont
  {Doyon}}, \ and\ \bibinfo {author} {\bibfnamefont {T.}~\bibnamefont
  {Yoshimura}},\ }\href@noop {} {\bibfield  {journal} {\bibinfo  {journal}
  {Phys. Rev. X}\ }\textbf {\bibinfo {volume} {6}},\ \bibinfo {pages} {041065}
  (\bibinfo {year} {2016})}\BibitemShut {NoStop}%
\bibitem [{\citenamefont {Bulchandani}\ \emph {et~al.}(2017)\citenamefont
  {Bulchandani}, \citenamefont {Vasseur}, \citenamefont {Karrasch},\ and\
  \citenamefont {Moore}}]{2017integrability2}%
  \BibitemOpen
  \bibfield  {author} {\bibinfo {author} {\bibfnamefont {V.~B.}\ \bibnamefont
  {Bulchandani}}, \bibinfo {author} {\bibfnamefont {R.}~\bibnamefont
  {Vasseur}}, \bibinfo {author} {\bibfnamefont {C.}~\bibnamefont {Karrasch}}, \
  and\ \bibinfo {author} {\bibfnamefont {J.~E.}\ \bibnamefont {Moore}},\
  }\href@noop {} {\bibfield  {journal} {\bibinfo  {journal} {Phys. Rev. Lett.}\
  }\textbf {\bibinfo {volume} {119}},\ \bibinfo {pages} {220604} (\bibinfo
  {year} {2017})}\BibitemShut {NoStop}%
\bibitem [{\citenamefont {Lucas}\ and\ \citenamefont
  {Hartnoll}(2017)}]{2017resistivitybound}%
  \BibitemOpen
  \bibfield  {author} {\bibinfo {author} {\bibfnamefont {A.}~\bibnamefont
  {Lucas}}\ and\ \bibinfo {author} {\bibfnamefont {S.~A.}\ \bibnamefont
  {Hartnoll}},\ }\href@noop {} {\bibfield  {journal} {\bibinfo  {journal}
  {Proceedings of the National Academy of Sciences}\ }\textbf {\bibinfo
  {volume} {114}},\ \bibinfo {pages} {11344} (\bibinfo {year}
  {2017})}\BibitemShut {NoStop}%
\bibitem [{\citenamefont {Scopelliti}\ \emph {et~al.}(2017)\citenamefont
  {Scopelliti}, \citenamefont {Schalm},\ and\ \citenamefont
  {Lucas}}]{2017curvedspace}%
  \BibitemOpen
  \bibfield  {author} {\bibinfo {author} {\bibfnamefont {V.}~\bibnamefont
  {Scopelliti}}, \bibinfo {author} {\bibfnamefont {K.}~\bibnamefont {Schalm}},
  \ and\ \bibinfo {author} {\bibfnamefont {A.}~\bibnamefont {Lucas}},\
  }\href@noop {} {\bibfield  {journal} {\bibinfo  {journal} {Phys. Rev. B}\
  }\textbf {\bibinfo {volume} {96}},\ \bibinfo {pages} {075150} (\bibinfo
  {year} {2017})}\BibitemShut {NoStop}%
\bibitem [{\citenamefont {Delacr\'etaz}\ \emph {et~al.}(2017)\citenamefont
  {Delacr\'etaz}, \citenamefont {Gout\'eraux}, \citenamefont {Hartnoll},\ and\
  \citenamefont {Karlsson}}]{2017lucadensitywave}%
  \BibitemOpen
  \bibfield  {author} {\bibinfo {author} {\bibfnamefont {L.~V.}\ \bibnamefont
  {Delacr\'etaz}}, \bibinfo {author} {\bibfnamefont {B.}~\bibnamefont
  {Gout\'eraux}}, \bibinfo {author} {\bibfnamefont {S.~A.}\ \bibnamefont
  {Hartnoll}}, \ and\ \bibinfo {author} {\bibfnamefont {A.}~\bibnamefont
  {Karlsson}},\ }\href@noop {} {\bibfield  {journal} {\bibinfo  {journal}
  {Phys. Rev. B}\ }\textbf {\bibinfo {volume} {96}},\ \bibinfo {pages} {195128}
  (\bibinfo {year} {2017})}\BibitemShut {NoStop}%
\bibitem [{\citenamefont {Delacr\'etaz}\ and\ \citenamefont
  {Gromov}(2017)}]{2017lucahallviscosity}%
  \BibitemOpen
  \bibfield  {author} {\bibinfo {author} {\bibfnamefont {L.~V.}\ \bibnamefont
  {Delacr\'etaz}}\ and\ \bibinfo {author} {\bibfnamefont {A.}~\bibnamefont
  {Gromov}},\ }\href@noop {} {\bibfield  {journal} {\bibinfo  {journal} {Phys.
  Rev. Lett.}\ }\textbf {\bibinfo {volume} {119}},\ \bibinfo {pages} {226602}
  (\bibinfo {year} {2017})}\BibitemShut {NoStop}%
\bibitem [{\citenamefont {Scaffidi}\ \emph {et~al.}(2017)\citenamefont
  {Scaffidi}, \citenamefont {Nandi}, \citenamefont {Schmidt}, \citenamefont
  {Mackenzie},\ and\ \citenamefont {Moore}}]{2017moorehydro}%
  \BibitemOpen
  \bibfield  {author} {\bibinfo {author} {\bibfnamefont {T.}~\bibnamefont
  {Scaffidi}}, \bibinfo {author} {\bibfnamefont {N.}~\bibnamefont {Nandi}},
  \bibinfo {author} {\bibfnamefont {B.}~\bibnamefont {Schmidt}}, \bibinfo
  {author} {\bibfnamefont {A.~P.}\ \bibnamefont {Mackenzie}}, \ and\ \bibinfo
  {author} {\bibfnamefont {J.~E.}\ \bibnamefont {Moore}},\ }\href@noop {}
  {\bibfield  {journal} {\bibinfo  {journal} {Phys. Rev. Lett.}\ }\textbf
  {\bibinfo {volume} {118}},\ \bibinfo {pages} {226601} (\bibinfo {year}
  {2017})}\BibitemShut {NoStop}%
\bibitem [{\citenamefont {Rondin}\ \emph {et~al.}(2014)\citenamefont {Rondin},
  \citenamefont {Tetienne}, \citenamefont {Hingant}, \citenamefont {Roch},
  \citenamefont {Maletinsky},\ and\ \citenamefont {Jacques}}]{2014nvreview}%
  \BibitemOpen
  \bibfield  {author} {\bibinfo {author} {\bibfnamefont {L.}~\bibnamefont
  {Rondin}}, \bibinfo {author} {\bibfnamefont {J.-P.}\ \bibnamefont
  {Tetienne}}, \bibinfo {author} {\bibfnamefont {T.}~\bibnamefont {Hingant}},
  \bibinfo {author} {\bibfnamefont {J.-F.}\ \bibnamefont {Roch}}, \bibinfo
  {author} {\bibfnamefont {P.}~\bibnamefont {Maletinsky}}, \ and\ \bibinfo
  {author} {\bibfnamefont {V.}~\bibnamefont {Jacques}},\ }\href@noop {}
  {\bibfield  {journal} {\bibinfo  {journal} {Reports on Progress in Physics}\
  }\textbf {\bibinfo {volume} {77}},\ \bibinfo {pages} {056503} (\bibinfo
  {year} {2014})}\BibitemShut {NoStop}%
\bibitem [{\citenamefont {Degen}\ \emph {et~al.}(2017)\citenamefont {Degen},
  \citenamefont {Reinhard},\ and\ \citenamefont {Cappellaro}}]{2017cappellaro}%
  \BibitemOpen
  \bibfield  {author} {\bibinfo {author} {\bibfnamefont {C.~L.}\ \bibnamefont
  {Degen}}, \bibinfo {author} {\bibfnamefont {F.}~\bibnamefont {Reinhard}}, \
  and\ \bibinfo {author} {\bibfnamefont {P.}~\bibnamefont {Cappellaro}},\
  }\href {\doibase 10.1103/RevModPhys.89.035002} {\bibfield  {journal}
  {\bibinfo  {journal} {Rev. Mod. Phys.}\ }\textbf {\bibinfo {volume} {89}},\
  \bibinfo {pages} {035002} (\bibinfo {year} {2017})}\BibitemShut {NoStop}%
\bibitem [{\citenamefont {Dyson}(1956)}]{1956dyson}%
  \BibitemOpen
  \bibfield  {author} {\bibinfo {author} {\bibfnamefont {F.~J.}\ \bibnamefont
  {Dyson}},\ }\href@noop {} {\bibfield  {journal} {\bibinfo  {journal} {Phys.
  Rev.}\ }\textbf {\bibinfo {volume} {102}},\ \bibinfo {pages} {1217} (\bibinfo
  {year} {1956})}\BibitemShut {NoStop}%
\bibitem [{\citenamefont {Lucas}\ and\ \citenamefont
  {Das~Sarma}(2018)}]{2018electronsound}%
  \BibitemOpen
  \bibfield  {author} {\bibinfo {author} {\bibfnamefont {A.}~\bibnamefont
  {Lucas}}\ and\ \bibinfo {author} {\bibfnamefont {S.}~\bibnamefont
  {Das~Sarma}},\ }\href@noop {} {\bibfield  {journal} {\bibinfo  {journal}
  {Phys. Rev. B}\ }\textbf {\bibinfo {volume} {97}},\ \bibinfo {pages} {115449}
  (\bibinfo {year} {2018})}\BibitemShut {NoStop}%
\bibitem [{\citenamefont {Iacocca}\ \emph {et~al.}(2017)\citenamefont
  {Iacocca}, \citenamefont {Silva},\ and\ \citenamefont
  {Hoefer}}]{2017galileansymmetrybreaking}%
  \BibitemOpen
  \bibfield  {author} {\bibinfo {author} {\bibfnamefont {E.}~\bibnamefont
  {Iacocca}}, \bibinfo {author} {\bibfnamefont {T.~J.}\ \bibnamefont {Silva}},
  \ and\ \bibinfo {author} {\bibfnamefont {M.~A.}\ \bibnamefont {Hoefer}},\
  }\href {\doibase 10.1103/PhysRevLett.118.017203} {\bibfield  {journal}
  {\bibinfo  {journal} {Phys. Rev. Lett.}\ }\textbf {\bibinfo {volume} {118}},\
  \bibinfo {pages} {017203} (\bibinfo {year} {2017})}\BibitemShut {NoStop}%
\bibitem [{\citenamefont {Demler}\ and\ \citenamefont
  {Maltsev}(2011)}]{2011demler}%
  \BibitemOpen
  \bibfield  {author} {\bibinfo {author} {\bibfnamefont {E.}~\bibnamefont
  {Demler}}\ and\ \bibinfo {author} {\bibfnamefont {A.}~\bibnamefont
  {Maltsev}},\ }\href {\doibase https://doi.org/10.1016/j.aop.2011.04.001}
  {\bibfield  {journal} {\bibinfo  {journal} {Annals of Physics}\ }\textbf
  {\bibinfo {volume} {326}},\ \bibinfo {pages} {1775 } (\bibinfo {year}
  {2011})},\ \bibinfo {note} {july 2011 Special Issue}\BibitemShut {NoStop}%
\bibitem [{\citenamefont {Demler}\ \emph {et~al.}(2017)\citenamefont {Demler},
  \citenamefont {Maltsev},\ and\ \citenamefont {Prokofiev}}]{2017demler}%
  \BibitemOpen
  \bibfield  {author} {\bibinfo {author} {\bibfnamefont {E.~A.}\ \bibnamefont
  {Demler}}, \bibinfo {author} {\bibfnamefont {A.~Y.}\ \bibnamefont {Maltsev}},
  \ and\ \bibinfo {author} {\bibfnamefont {A.~O.}\ \bibnamefont {Prokofiev}},\
  }\href@noop {} {\bibfield  {journal} {\bibinfo  {journal} {J. Phys. B}\
  }\textbf {\bibinfo {volume} {50}},\ \bibinfo {pages} {124001} (\bibinfo
  {year} {2017})}\BibitemShut {NoStop}%
\bibitem [{\citenamefont {Halperin}\ and\ \citenamefont
  {Hohenberg}(1969)}]{1969halperinhohenberg}%
  \BibitemOpen
  \bibfield  {author} {\bibinfo {author} {\bibfnamefont {B.~I.}\ \bibnamefont
  {Halperin}}\ and\ \bibinfo {author} {\bibfnamefont {P.~C.}\ \bibnamefont
  {Hohenberg}},\ }\href@noop {} {\bibfield  {journal} {\bibinfo  {journal}
  {Phys. Rev.}\ }\textbf {\bibinfo {volume} {188}},\ \bibinfo {pages} {898}
  (\bibinfo {year} {1969})}\BibitemShut {NoStop}%
\bibitem [{\citenamefont {Reiter}(1968)}]{1968reiter}%
  \BibitemOpen
  \bibfield  {author} {\bibinfo {author} {\bibfnamefont {G.~F.}\ \bibnamefont
  {Reiter}},\ }\href@noop {} {\bibfield  {journal} {\bibinfo  {journal} {Phys.
  Rev.}\ }\textbf {\bibinfo {volume} {175}},\ \bibinfo {pages} {631} (\bibinfo
  {year} {1968})}\BibitemShut {NoStop}%
\bibitem [{\citenamefont {Michel}\ and\ \citenamefont
  {Schwabl}(1969)}]{1969michel}%
  \BibitemOpen
  \bibfield  {author} {\bibinfo {author} {\bibfnamefont {K.}~\bibnamefont
  {Michel}}\ and\ \bibinfo {author} {\bibfnamefont {F.}~\bibnamefont
  {Schwabl}},\ }\href@noop {} {\bibfield  {journal} {\bibinfo  {journal} {Solid
  State Communications}\ }\textbf {\bibinfo {volume} {7}},\ \bibinfo {pages}
  {1781} (\bibinfo {year} {1969})}\BibitemShut {NoStop}%
\bibitem [{\citenamefont {Schwabl}\ and\ \citenamefont
  {Michel}(1970)}]{1970Schwabl}%
  \BibitemOpen
  \bibfield  {author} {\bibinfo {author} {\bibfnamefont {F.}~\bibnamefont
  {Schwabl}}\ and\ \bibinfo {author} {\bibfnamefont {K.~H.}\ \bibnamefont
  {Michel}},\ }\href@noop {} {\bibfield  {journal} {\bibinfo  {journal} {Phys.
  Rev. B}\ }\textbf {\bibinfo {volume} {2}},\ \bibinfo {pages} {189} (\bibinfo
  {year} {1970})}\BibitemShut {NoStop}%
\bibitem [{\citenamefont {van~der Sar}\ \emph {et~al.}(2015)\citenamefont
  {van~der Sar}, \citenamefont {Casola}, \citenamefont {Walsworth},\ and\
  \citenamefont {Yacoby}}]{2015nvmagnons}%
  \BibitemOpen
  \bibfield  {author} {\bibinfo {author} {\bibfnamefont {T.}~\bibnamefont
  {van~der Sar}}, \bibinfo {author} {\bibfnamefont {F.}~\bibnamefont {Casola}},
  \bibinfo {author} {\bibfnamefont {R.}~\bibnamefont {Walsworth}}, \ and\
  \bibinfo {author} {\bibfnamefont {A.}~\bibnamefont {Yacoby}},\ }\href@noop {}
  {\bibfield  {journal} {\bibinfo  {journal} {Nat Commun}\ } (\bibinfo {year}
  {2015})}\BibitemShut {NoStop}%
\bibitem [{\citenamefont {Du}\ \emph {et~al.}(2017)\citenamefont {Du},
  \citenamefont {van~der Sar}, \citenamefont {Zhou}, \citenamefont {Upadhyaya},
  \citenamefont {Casola}, \citenamefont {Zhang}, \citenamefont {Onbasli},
  \citenamefont {Ross}, \citenamefont {Walsworth}, \citenamefont
  {Tserkovnyak},\ and\ \citenamefont {Yacoby}}]{2017du}%
  \BibitemOpen
  \bibfield  {author} {\bibinfo {author} {\bibfnamefont {C.}~\bibnamefont
  {Du}}, \bibinfo {author} {\bibfnamefont {T.}~\bibnamefont {van~der Sar}},
  \bibinfo {author} {\bibfnamefont {T.~X.}\ \bibnamefont {Zhou}}, \bibinfo
  {author} {\bibfnamefont {P.}~\bibnamefont {Upadhyaya}}, \bibinfo {author}
  {\bibfnamefont {F.}~\bibnamefont {Casola}}, \bibinfo {author} {\bibfnamefont
  {H.}~\bibnamefont {Zhang}}, \bibinfo {author} {\bibfnamefont {M.~C.}\
  \bibnamefont {Onbasli}}, \bibinfo {author} {\bibfnamefont {C.~A.}\
  \bibnamefont {Ross}}, \bibinfo {author} {\bibfnamefont {R.~L.}\ \bibnamefont
  {Walsworth}}, \bibinfo {author} {\bibfnamefont {Y.}~\bibnamefont
  {Tserkovnyak}}, \ and\ \bibinfo {author} {\bibfnamefont {A.}~\bibnamefont
  {Yacoby}},\ }\href@noop {} {\bibfield  {journal} {\bibinfo  {journal}
  {Science}\ }\textbf {\bibinfo {volume} {357}},\ \bibinfo {pages} {195}
  (\bibinfo {year} {2017})}\BibitemShut {NoStop}%
\bibitem [{\citenamefont {Grinolds}\ \emph {et~al.}(2013)\citenamefont
  {Grinolds}, \citenamefont {Hong}, \citenamefont {Maletinsky}, \citenamefont
  {Luan}, \citenamefont {Lukin}, \citenamefont {Walsworth},\ and\ \citenamefont
  {Yacoby}}]{2013singlespinimaging}%
  \BibitemOpen
  \bibfield  {author} {\bibinfo {author} {\bibfnamefont {M.~S.}\ \bibnamefont
  {Grinolds}}, \bibinfo {author} {\bibfnamefont {S.}~\bibnamefont {Hong}},
  \bibinfo {author} {\bibfnamefont {P.}~\bibnamefont {Maletinsky}}, \bibinfo
  {author} {\bibfnamefont {L.}~\bibnamefont {Luan}}, \bibinfo {author}
  {\bibfnamefont {M.~D.}\ \bibnamefont {Lukin}}, \bibinfo {author}
  {\bibfnamefont {R.~L.}\ \bibnamefont {Walsworth}}, \ and\ \bibinfo {author}
  {\bibfnamefont {A.}~\bibnamefont {Yacoby}},\ }\href@noop {} {\bibfield
  {journal} {\bibinfo  {journal} {Nat Phys}\ }\textbf {\bibinfo {volume} {9}},\
  \bibinfo {pages} {215} (\bibinfo {year} {2013})}\BibitemShut {NoStop}%
\bibitem [{\citenamefont {Tetienne}\ \emph {et~al.}(2015)\citenamefont
  {Tetienne}, \citenamefont {Hingant}, \citenamefont {Martínez}, \citenamefont
  {Rohart}, \citenamefont {Thiaville}, \citenamefont {Diez}, \citenamefont
  {Garcia}, \citenamefont {Adam}, \citenamefont {Kim}, \citenamefont {Roch},
  \citenamefont {Miron}, \citenamefont {Gaudin}, \citenamefont {Vila},
  \citenamefont {Ocker}, \citenamefont {Ravelosona},\ and\ \citenamefont
  {Jacques}}]{2016nvdomainwalls}%
  \BibitemOpen
  \bibfield  {author} {\bibinfo {author} {\bibfnamefont {J.-P.}\ \bibnamefont
  {Tetienne}}, \bibinfo {author} {\bibfnamefont {T.}~\bibnamefont {Hingant}},
  \bibinfo {author} {\bibfnamefont {L.}~\bibnamefont {Martínez}}, \bibinfo
  {author} {\bibfnamefont {S.}~\bibnamefont {Rohart}}, \bibinfo {author}
  {\bibfnamefont {A.}~\bibnamefont {Thiaville}}, \bibinfo {author}
  {\bibfnamefont {L.~H.}\ \bibnamefont {Diez}}, \bibinfo {author}
  {\bibfnamefont {K.}~\bibnamefont {Garcia}}, \bibinfo {author} {\bibfnamefont
  {J.-P.}\ \bibnamefont {Adam}}, \bibinfo {author} {\bibfnamefont {J.-V.}\
  \bibnamefont {Kim}}, \bibinfo {author} {\bibfnamefont {J.-F.}\ \bibnamefont
  {Roch}}, \bibinfo {author} {\bibfnamefont {I.}~\bibnamefont {Miron}},
  \bibinfo {author} {\bibfnamefont {G.}~\bibnamefont {Gaudin}}, \bibinfo
  {author} {\bibfnamefont {L.}~\bibnamefont {Vila}}, \bibinfo {author}
  {\bibfnamefont {B.}~\bibnamefont {Ocker}}, \bibinfo {author} {\bibfnamefont
  {D.}~\bibnamefont {Ravelosona}}, \ and\ \bibinfo {author} {\bibfnamefont
  {V.}~\bibnamefont {Jacques}},\ }\href@noop {} {\bibfield  {journal} {\bibinfo
   {journal} {Nat Commun}\ } (\bibinfo {year} {2015})}\BibitemShut {NoStop}%
\bibitem [{\citenamefont {Kolkowitz}\ \emph {et~al.}(2015)\citenamefont
  {Kolkowitz}, \citenamefont {Safira}, \citenamefont {High}, \citenamefont
  {Devlin}, \citenamefont {Choi}, \citenamefont {Unterreithmeier},
  \citenamefont {Patterson}, \citenamefont {Zibrov}, \citenamefont
  {Manucharyan}, \citenamefont {Park},\ and\ \citenamefont
  {Lukin}}]{2015kolkowitz}%
  \BibitemOpen
  \bibfield  {author} {\bibinfo {author} {\bibfnamefont {S.}~\bibnamefont
  {Kolkowitz}}, \bibinfo {author} {\bibfnamefont {A.}~\bibnamefont {Safira}},
  \bibinfo {author} {\bibfnamefont {A.~A.}\ \bibnamefont {High}}, \bibinfo
  {author} {\bibfnamefont {R.~C.}\ \bibnamefont {Devlin}}, \bibinfo {author}
  {\bibfnamefont {S.}~\bibnamefont {Choi}}, \bibinfo {author} {\bibfnamefont
  {Q.~P.}\ \bibnamefont {Unterreithmeier}}, \bibinfo {author} {\bibfnamefont
  {D.}~\bibnamefont {Patterson}}, \bibinfo {author} {\bibfnamefont {A.~S.}\
  \bibnamefont {Zibrov}}, \bibinfo {author} {\bibfnamefont {V.~E.}\
  \bibnamefont {Manucharyan}}, \bibinfo {author} {\bibfnamefont
  {H.}~\bibnamefont {Park}}, \ and\ \bibinfo {author} {\bibfnamefont {M.~D.}\
  \bibnamefont {Lukin}},\ }\href@noop {} {\bibfield  {journal} {\bibinfo
  {journal} {Science}\ }\textbf {\bibinfo {volume} {347}},\ \bibinfo {pages}
  {1129} (\bibinfo {year} {2015})}\BibitemShut {NoStop}%
\bibitem [{\citenamefont {Agarwal}\ \emph {et~al.}(2017)\citenamefont
  {Agarwal}, \citenamefont {Schmidt}, \citenamefont {Halperin}, \citenamefont
  {Oganesyan}, \citenamefont {Zar\'and}, \citenamefont {Lukin},\ and\
  \citenamefont {Demler}}]{2017kartieknoise}%
  \BibitemOpen
  \bibfield  {author} {\bibinfo {author} {\bibfnamefont {K.}~\bibnamefont
  {Agarwal}}, \bibinfo {author} {\bibfnamefont {R.}~\bibnamefont {Schmidt}},
  \bibinfo {author} {\bibfnamefont {B.}~\bibnamefont {Halperin}}, \bibinfo
  {author} {\bibfnamefont {V.}~\bibnamefont {Oganesyan}}, \bibinfo {author}
  {\bibfnamefont {G.}~\bibnamefont {Zar\'and}}, \bibinfo {author}
  {\bibfnamefont {M.~D.}\ \bibnamefont {Lukin}}, \ and\ \bibinfo {author}
  {\bibfnamefont {E.}~\bibnamefont {Demler}},\ }\href@noop {} {\bibfield
  {journal} {\bibinfo  {journal} {Phys. Rev. B}\ }\textbf {\bibinfo {volume}
  {95}},\ \bibinfo {pages} {155107} (\bibinfo {year} {2017})}\BibitemShut
  {NoStop}%
\bibitem [{\citenamefont {{Rodriguez-Nieva}}\ \emph {et~al.}(2018)\citenamefont
  {{Rodriguez-Nieva}}, \citenamefont {{Agarwal}}, \citenamefont {{Giamarchi}},
  \citenamefont {{Halperin}}, \citenamefont {{Lukin}},\ and\ \citenamefont
  {{Demler}}}]{2018nv-wire}%
  \BibitemOpen
  \bibfield  {author} {\bibinfo {author} {\bibfnamefont {J.~F.}\ \bibnamefont
  {{Rodriguez-Nieva}}}, \bibinfo {author} {\bibfnamefont {K.}~\bibnamefont
  {{Agarwal}}}, \bibinfo {author} {\bibfnamefont {T.}~\bibnamefont
  {{Giamarchi}}}, \bibinfo {author} {\bibfnamefont {B.~I.}\ \bibnamefont
  {{Halperin}}}, \bibinfo {author} {\bibfnamefont {M.~D.}\ \bibnamefont
  {{Lukin}}}, \ and\ \bibinfo {author} {\bibfnamefont {E.}~\bibnamefont
  {{Demler}}},\ }\href@noop {} {\bibfield  {journal} {\bibinfo  {journal}
  {ArXiv e-prints}\ } (\bibinfo {year} {2018})},\ \Eprint
  {http://arxiv.org/abs/1803.01521} {arXiv:1803.01521} \BibitemShut {NoStop}%
\bibitem [{\citenamefont {Flebus}\ and\ \citenamefont
  {Tserkovnyak}(2018)}]{2018flebus}%
  \BibitemOpen
  \bibfield  {author} {\bibinfo {author} {\bibfnamefont {B.}~\bibnamefont
  {Flebus}}\ and\ \bibinfo {author} {\bibfnamefont {Y.}~\bibnamefont
  {Tserkovnyak}},\ }\href@noop {} {\bibfield  {journal} {\bibinfo  {journal}
  {ArXiv e-prints}\ } (\bibinfo {year} {2018})},\ \Eprint
  {http://arxiv.org/abs/1804.02417} {arXiv:1804.02417} \BibitemShut {NoStop}%
\bibitem [{\citenamefont {{Chatterjee}}\ \emph {et~al.}(2018)\citenamefont
  {{Chatterjee}}, \citenamefont {{Rodriguez-Nieva}},\ and\ \citenamefont
  {{Demler}}}]{2018nv-sl}%
  \BibitemOpen
  \bibfield  {author} {\bibinfo {author} {\bibfnamefont {S.}~\bibnamefont
  {{Chatterjee}}}, \bibinfo {author} {\bibfnamefont {J.~F.}\ \bibnamefont
  {{Rodriguez-Nieva}}}, \ and\ \bibinfo {author} {\bibfnamefont
  {E.}~\bibnamefont {{Demler}}},\ }\href@noop {} {\bibfield  {journal}
  {\bibinfo  {journal} {ArXiv e-prints}\ } (\bibinfo {year} {2018})},\ \Eprint
  {http://arxiv.org/abs/1810.04183} {arXiv:1810.04183} \BibitemShut {NoStop}%
\bibitem [{two()}]{twomagnons}%
  \BibitemOpen
  \href@noop {} {}\bibinfo {note} {In fact, not only is $|{\bm k},{\bm
  p}\rangle = S_{\bm k}^+S_{\bm p}^+|{\rm F}\rangle$ not diagonal, but they are
  not properly normalized nor do they form an orthogonal basis, see discussion
  in Supplement.}\BibitemShut {Stop}%
\bibitem [{\citenamefont {Mattis}(2006)}]{mattisbook}%
  \BibitemOpen
  \bibfield  {author} {\bibinfo {author} {\bibfnamefont {D.~C.}\ \bibnamefont
  {Mattis}},\ }\href@noop {} {\emph {\bibinfo {title} {The Theory of Magnetism
  Made Simple}}}\ (\bibinfo  {publisher} {World Scientific},\ \bibinfo {year}
  {2006})\BibitemShut {NoStop}%
\bibitem [{sup()}]{supplement}%
  \BibitemOpen
  \href@noop {} {}\bibinfo {note} {Supplementary information}\BibitemShut
  {NoStop}%
\bibitem [{\citenamefont {Flebus}\ \emph {et~al.}(2016)\citenamefont {Flebus},
  \citenamefont {Bender}, \citenamefont {Tserkovnyak},\ and\ \citenamefont
  {Duine}}]{2016flebus}%
  \BibitemOpen
  \bibfield  {author} {\bibinfo {author} {\bibfnamefont {B.}~\bibnamefont
  {Flebus}}, \bibinfo {author} {\bibfnamefont {S.~A.}\ \bibnamefont {Bender}},
  \bibinfo {author} {\bibfnamefont {Y.}~\bibnamefont {Tserkovnyak}}, \ and\
  \bibinfo {author} {\bibfnamefont {R.~A.}\ \bibnamefont {Duine}},\ }\href
  {\doibase 10.1103/PhysRevLett.116.117201} {\bibfield  {journal} {\bibinfo
  {journal} {Phys. Rev. Lett.}\ }\textbf {\bibinfo {volume} {116}},\ \bibinfo
  {pages} {117201} (\bibinfo {year} {2016})}\BibitemShut {NoStop}%
\end{thebibliography}
\end{document}